\documentclass{getwriting}
\usepackage{amsmath,amssymb}
\usepackage{bm}
\usepackage{ulem}
\usepackage[numbers]{natbib}	
\usepackage{graphicx}

\title{Waveform distortion for temperature compensation and synchronization in circadian rhythms: \\An approach based on the renormalization group method}
\author[1]{Shingo Gibo \thanks{shingo.gibo@riken.jp}}
\author[2]{Teiji Kunihiro}
\author[1]{Tetsuo Hatsuda}
\author[1]{Gen Kurosawa \thanks{g.kurosawa@riken.jp}}
\affil[1]{\footnotesize Interdisciplinary Theoretical and Mathematical Sciences Program (iTHEMS), RIKEN}
\affil[2]{Yukawa Institute for Theoretical Physics (YITP), Kyoto University}
\begin{document}
\maketitle
\begin{abstract}
Numerous biological processes accelerate as temperatures increase, but the period of circadian rhythms remains constant, known as temperature compensation, while synchronizing with the 24h light-dark cycle. 
We
theoretically explores 
the possible relevance of waveform distortions in circadian gene-protein dynamics to
the temperature compensation and synchronization.
Our analysis of the Goodwin model provides a
coherent explanation 
of most of temperature compensation hypotheses. Using the renormalization group 
method, we analytically
demonstrate that the decreasing phase of circadian protein oscillations should lengthen with increasing temperature, 
leading to waveform distortions to maintain a stable period. This waveform-period correlation also occurs in other oscillators like Lotka-Volterra and van der Pol models. A reanalysis of known data nicely confirms
our findings on waveform distortion and its impact on synchronization range. 
Thus we conclude that circadian rhythm waveforms are fundamental to both temperature compensation and synchronization. 

\textbf{Keywords:} waveform distortion, renormalization group method, circadian rhythms, temperature compensation, synchronization

\end{abstract}

\section*{Author Summary}
Our daily rhythms are underlain by gene regulatory and biochemical networks, called circadian clocks. Although most biochemical reactions accelerate as temperature increases, the period of circadian rhythms is almost constant even with increasing temperature. This phenomenon is called temperature compensation, and the mechanism is still unclear. By applying a method of theoretical physics, the renormalization group method to a biological problem for the first time, we revealed that the waveform of gene dynamics should be more distorted from sinusoidal wave at higher temperature when the circadian period is stable to changes in temperature. This prediction as for the importance of waveform in temperature compensation is verified by analyzing published experimental data. Notably, the correlation between period and waveform distortion holds for other oscillator models, indicating the waveform distortion is important for determining the period in various types of oscillatory systems. Another unsolved problem of circadian clocks is to synchronize with environmental light-dark cycles. By theoretically analyzing a circadian clock model, we found that the frequency range for synchronization becomes narrower when the waveform is distorted.

\section{Introduction}
Humans exhibit sleep-wake cycles with an approximate 24h period, and these cycles persist 
under constant environmental conditions, a phenomenon termed the circadian rhythm. This temporal regulation 
exists in both humans and various organisms such as molds, plants, and insects \citep{dunl99, harmer01, edgar12}. Recent advances in genetic research 
through  insects, molds, mammals, and plants have unveiled 
that genes and proteins 
are involved as integral components in the primary mechanism governing autonomous circadian rhythms~\citep{dunl99, ha90, ko06, sato06}. 

Understanding circadian rhythms holds promise for deciphering a multitude of sleep patterns, including sleep disorders such as advanced sleep phase syndrome (characterized by early awakening around 4:00 am), delayed sleep phase syndrome (marked by late awakening), non-24h sleep-wake disorder, and narcolepsy \cite{duffy05}. Notably, advanced and delayed sleep phase syndromes are believed to be linked to the circadian rhythm period \cite{toh01, vanselow06, xu07, brown08, granada13, fustin18}. Ongoing studies explore possible correlations between genetic characteristics revealed by large-scale genetic analysis and various sleep patterns \cite{hu16, lane16, shi17}. However, the nature of the system is so intricate that it remains a challenge to link sleep patterns to specific genes. In such a situation, it would be meaningful to have recourse to mathematical models and obtain possible hints for the linkage and hopefully suggestions for studies of genetic dynamics.

One unresolved fundamental issue in circadian rhythm research is temperature
compensation \cite{ha57, ruoff92, rub99, hatakeyama15, zhou15, kuro17, sasai22, schmal23, chakra23} 
, in which the period 
keeps constant despite temperature-induced changes in reaction rates. 
Despite the extensive experimental and theoretical research on 
temperature compensation, the mechanism has remained elusive. 
Hypotheses have been proposed to explain temperature compensation, 
including the balance hypothesis, critical-reaction hypothesis, 
temperature-amplitude coupling hypothesis, and waveform hypothesis. 
The balance hypothesis proposes that the stability of the circadian period 
with temperature arises from a balance between period-lengthening and 
period-shortening reactions \cite{ha57, ruoff92, kuro05}. 
The critical-reaction hypothesis assumes that there should be critical 
reactions that determine the circadian period. If these reaction 
rates are stable against temperature variations, 
then the circadian period will similarly remain stable
\citep{tera07, hong09, iso09}. The temperature-amplitude 
coupling hypothesis suggests that temperature-sensitive amplitudes 
in gene activity rhythms should generate a stable period by generating 
larger amplitudes at higher temperatures \cite{kuro17, lak91}. 
Lastly, the waveform hypothesis proposes that temperature-sensitive 
waveforms in gene activity rhythms should be correlated with a stable 
period in a manner that their higher harmonic components become 
larger and the distortion of the waveform increases at higher
temperatures \cite{kuro19}.  

Another unresolved issue in circadian rhythm research is synchronization 
with 24h environmental light-dark cycles.
Previous theoretical and experimental studies on synchronization revealed 
that if the internal period of the oscillation closely matches the 
external period, then it is more likely to synchronize with the forced 
period \cite{steve15, pikovsky03, abra10, granada11}. 
Additionally, experimental studies on several species uncovered genes and 
proteins in circadian systems affected by a light pulse \cite{crosthwaite95, shigeyoshi97}. 
In reality, the circadian rhythm must adjust to the 24h light-dark cycle while 
maintaining a temperature-compensated circadian period. 
Therefore, multiple questions arise. (i) Given the significant temperature 
variations between seasons, how do organisms synchronize their circadian 
rhythms with the 24h light-dark cycle across various temperatures 
\cite{hatakeyama15, abra10, take07}? (ii) if the gene activity 
rhythm of the circadian rhythms becomes more distorted as temperatures increase 
to achieve temperature compensation, how does the ease of synchronization 
change with temperature variations?
Theoretical analyses incorporating the findings of light pulse experiments
might provide further insights into these questions.
 
In the present paper, we investigate possible roles of the waveform distortion 
in temperature compensation based on analytical and numerical analyses 
of the Goodwin model for circadian rhythms and clarify how the 
waveform in gene activity rhythms tends to be more distorted at higher 
temperatures (e.g., steeper rise, longer tail) for temperature compensation.
To this end, we employ the renormalization group (RG) method, a powerful 
tool for analyzing various non-linear systems described by ordinary and partial 
differential equations, to derive global solutions that are valid in a global 
time domain \cite{golde89, chen94, kuni95, gra96, ku97, kuni97, sas97, mar99, vey07, dev08, chiba09, omal10, oono12, holz14, ei20, kuni22}. 

Combining an index for waveform distortion, namely non-sinusoidal 
power ($NS$) introduced by two of the present authors (KG) \citep{kuro19}, with the result of 
the RG method,
we can obtain both a unified picture of 
the above mentioned theoretical hypotheses 
(balance hypothesis, critical-reaction 
hypothesis, temperature-amplitude coupling hypothesis, and waveform hypothesis) 
and quantify previous experimental data on Drosophila \citep{kidd15}. 
 Our analyses demonstrate that the fundamental role of the waveform distortions 
 in temperature compensation from both theoretical and experimental 
 perspectives in accordance with 
 the previous findings \citep{kuro19}.
Moreover, we reveal for the first time the mechanism by which 
the synchronization of circadian rhythms changes with temperature if 
the waveform in gene activity rhythms is more distorted at higher temperatures. 
We theoretically prove that the frequency range of the external force 
that synchronizes circadian rhythms becomes narrower if the waveform 
of gene activity rhythms is more distorted. This indicates that it is more 
difficult to synchronize with light-dark cycles at higher temperatures. The 
present result of synchronization is consistent with the previous experimental 
and numerical studies demonstrating that the magnitude of the phase shift 
caused by light pulses was smaller at higher temperature \cite{kuro19, nak80, varma13}.

\section*{Results}
\section{Waveform distortion in circadian rhythms}
\subsection{Index for waveform distortion}

Let the time dependence of a certain variable in the circadian rhythm system be 
expressed in a Fourier series as 
$ x(t)=\sum_{j=-\infty}^{\infty}a_{j}exp(i(\frac{2\pi}{\tau})jt))$, 
with $a_{j}$ being the Fourier coefficients of
the oscillatory time series. 
Then, we introduce an index for describing   
the distortion of $ x(t) $ from a sinusoidal shape as
\begin{equation}
NS=\left[ \frac{\sum_{j=1}^{\infty}|a_{j}|^{2}j^{m}}
{\sum_{j=1}^{\infty}|a_{j}|^{2}j^{q}} \right]^{\frac{1}{2}} (m>q\geq 0),
\label{eq1}
\end{equation}
where $m$ and $q$ are integers. 
Termed the "non-sinusoidal power ($NS$)", 
this index is designed to emphasize higher harmonics ($m>q$) 
as discussed in a previous paper~\citep{kuro19}. 
For instance, we have $NS=1$ when the time series has a sinusoidal waveform 
in which only the coefficients for the fundamental component are non-zero 
($a_{\pm 1}\not=0$). Conversely, for non-sinusoidal time series, 
the coefficients for higher harmonics are non-vanishing, resulting in $NS>1$. 
The previous theoretical work demonstrated that a more distorted waveform 
(larger $NS$) at higher temperature is necessary for temperature compensation 
in the four-variable negative-feedback model \cite{kuro19}. 
However, it is unclear whether the relevance of waveforms 
to temperature compensation 
found in previous research has general validity not 
restricted to some specific model.

\begin{figure}[t]	
 \begin{center}
  \includegraphics[width=16cm]{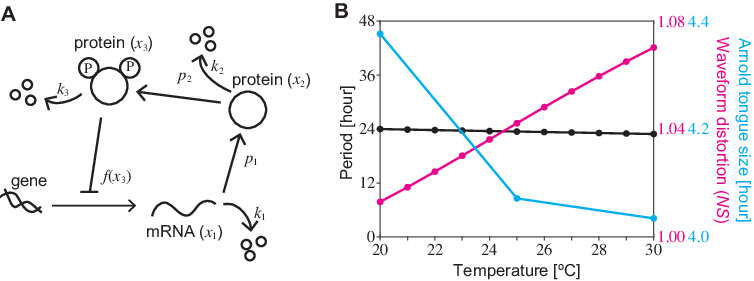}	
  \caption{(A) The circadian clock model and (B) the relevance of waveform characteristics to temperature compensation (\textit{magenta}) and synchronization (\textit{cyan}). The transcriptional-translational function is defined as $f(x_{3})=r/(1+(x_{3}/K)^{n})$, with the values of $K$ and $n$ set to be $0.0184$ and $11$, respectively. Assuming that the parameters $k_{1}$, $k_{2}$, $k_{3}$, $p_{1}$, $p_{2}$, and $r$ increase with increasing temperature, we model the reactions rates using the Arrhenius law $b_{i}=A_{i}\exp (-E_{i}/RT)$, where $b_{i}$, $E_{i}$, $A_{i}$, $R$, and $T$ are the rate constants, activation energies, frequency factors, gas constant ($R=8.314$), and absolute temperature, respectively. The values of $E_{i}$ and $A_{i}$ are detailed in Supplementary Table \ref{tabS1}. The synchronization region depends on the period of light-dark cycles and light intensity, known as the Arnold tongue \cite{granada11}. The size the Arnold tongue is defined by the difference between the lower limit of synchronization, $2\pi/\Omega_{\rm large}$,
   and the  upper limit of synchronization,  $2\pi/\Omega_{\rm small}$.} 
  When the period is constant 
  in a different temperature, the waveform distortion ($NS$) becomes larger, 
  and the synchronization range (size of the Arnold tongue) becomes narrower with increasing temperature as shown in (B) in the case of the  light intensity, $I=0.001$.
 \label{figure3}	
 \end{center}
\end{figure}

To explore the possible relevance of the waveform characteristics to temperature 
compensation and synchronization, we consider
the simplest model for circadian rhythms, known as the Goodwin model (Fig. \ref{figure3}A)
\cite{goo65}. This model incorporates negative-feedback regulation
of gene expression, a mechanism established as essential for 
transcriptional-translational oscillations. 
The three-component Goodwin model reads:
\begin{align}
&\frac{dx_{1}}{dt}=f(x_{3})-k_{1}x_{1},
\label{eq:G1} \\
&\frac{dx_{2}}{dt}=p_{1}x_{1}-k_{2}x_{2},
\label{eq:G2} \\
&\frac{dx_{3}}{dt}=p_{2}x_{2}-k_{3}x_{3},
\label{eq:G3}
\end{align}
where $x_{1}(t)$ represents mRNA abundance, and $x_{2}(t)$ 
and $x_{3}(t)$ denote protein abundance. 
The function $f(x_{3})$ in the model signifies transcriptional regulation, 
and the parameters $p_{1}$ and $p_{2}$ denote protein synthesis and 
phosphorylation rates, respectively,
and $k_{i}$ ($i=1,\,2,\,3$) represent degradation rates (Fig. \ref{figure3}A). 
By applying signal processing methods, Forger derived the period of this model 
as follows \cite{forg11}: 
\begin{equation}
\tau=\frac{2\pi}{\sqrt{k_{1}k_{2}+k_{2}k_{3}+k_{3}k_{1}}}
\left[ \frac{\sum_{j=1}^{\infty}|a_{j}|^{2}j^{4}}
{\sum_{j=1}^{\infty}|a_{j}|^{2}j^{2}} \right]^{\frac{1}{2}}.
\label{eq:tau}
\end{equation}
Subsequently, two of the present authors indicated that this formula implies 
that temperature compensation of the period in this model occurs 
only when the waveform ($NS$) is distorted as temperature 
increases \cite{kuro19}.
Suppose that all reactions become faster as temperature increases in the model. 
Then, one can numerically demonstrate that the waveform tends to be more 
distorted (larger $NS$) at higher temperatures for temperature compensation 
(Fig. \ref{figure3}B, magenta line). 
We call this mechanism the waveform hypothesis for temperature compensation. 

\subsection{Theories of temperature compensation}
The waveform hypothesis and the other three hypotheses for temperature 
compensation (balance hypothesis I, critical-reaction hypothesis II, and 
amplitude hypothesis III) are not mutually exclusive, which can be understood in a 
unified way through Eq. \eqref{eq:tau}:
\begin{description}
 \item{I.}
The balance hypothesis, previously explored theoretically by Ruoff \citep{ruoff92}, 
suggests that the temperature compensation of the period is caused by a balance 
between the effects of reactions that shorten the period and those that lengthen 
the period. Equation \eqref{eq:tau} illustrates that the balance between the effect 
of shortening the period and that of lengthening the period can be caused by 
changing the distortion of the waveform. 
\item{II.}
Equation \eqref{eq:tau} demonstrates that even if some of the governing 
reactions in circadian rhythms
are temperature-insensitive \cite{tera07, iso09}
 as discussed in the critical-reaction hypothesis. 
 Still, the temperature compensation  requires waveform distortion at high 
 temperatures 
 if reactions other than some of the governing reactions accelerate at higher 
 temperatures.
 \item{III.}
Our numerical analysis of the circadian model (Fig. \ref{figure3}B) 
show that when temperature compensation occurs, the waveform is more 
distorted at higher temperatures, 
and the amplitude of the oscillation is larger. This tendency for the amplitude to 
increase at high temperatures 
is consistent with the amplitude hypothesis. According to Eq. 
\eqref{eq:tau}, a greater distortion of the 
waveform at higher temperatures is necessary, but not sufficient, for 
temperature compensation. 
\end{description}

\subsection{Synchronization in circadian rhythms}
If the waveform is more distorted at higher temperatures for temperature 
compensation, then it would be intriguing to explore whether the temperature-
dependent waveform also affects the synchronization of circadian rhythms with 
environmental light-dark cycles at various temperatures. 
Theoretical and experimental studies of synchronization 
and circadian rhythms illustrated that the oscillation is more likely to 
synchronize with forcing cycles if the internal period is sufficiently 
close to the period of the forcing cycles \cite{steve15, pikovsky03, abra10, granada11}. In mammals and Neurospora, a light pulse is 
known to increase Per1 and frq mRNA expression \citep{crosthwaite95, shigeyoshi97}. To incorporate gene activation during the light phase, 
we employ
a model in which
Eq. \eqref{eq:G1} for the change in mRNA expression is modified to
\begin{equation}
 \frac{dx_{1}}{dt}=f(x_{3})-k_{1}x_{1}+I\cos(\Omega t),
\label{eq:G1-forced}
\end{equation}
where $I$ represents the light intensity and $\Omega$ is 
the angular frequency of the light-dark cycles. 

\subsection{Main results}
In Sections 3 and 4, we provide detailed discussions on how the waveform 
in gene activity rhythms should be distorted at higher temperatures using the 
RG method, as well as the synchronization in circadian rhythms.
The main results are pictorially summarized in Fig. \ref{figure3}B. 
The magenta line indicates that the waveform of the gene activity rhythms 
should be more distorted 
at higher temperatures for temperature compensation, whereas the cyan line 
indicates that synchronization 
with the light-dark cycles should become more difficult at higher temperatures 
because of the larger waveform distortion.

\section{Waveform distortion and temperature compensation}
\subsection{Numerical simulation of the waveform-period correlation}
Equation \eqref{eq:tau} and numerical simulations indicate that $NS$ tends 
to be larger when the period is relatively stable even with 
increased parameter values (see Fig. \ref{figure3}B). 
To quantitatively reveal the correlation between waveform and period, 
we conduct
numerical simulations using a circadian clock model. 
In the analysis of the circadian clock model, the transcription function 
$f(x_{3})=r/x_{3}^{n}$ was considered for simplicity. 
We first search
for parameter sets in which oscillations occur.
We define 
those parameter sets as the reference parameter sets. Because 
many biochemical parameters have not yet been measured, we prepared
100 random reference parameter sets 
for the oscillations. $k_{1}$, $k_{2}$, $k_{3}$, $p_{1}$, $p_{2}$, and $r$ 
were assigned uniformly distributed random values ranging from $0$ to $10$, 
and $n$ 
was assigned a uniformly distributed random 
integer ranging from $9$ to $15$. The period obtained with each reference 
parameter set 
was denoted as $\tau_1$.
Next, reaction rates often follow the Arrhenius equation, which states that a $10$°C 
rise in temperature increases 
the reaction rate by a factor of $2$-$3$. To incorporate the effect of high 
temperature, instead of 
using the Arrhenius equation, each parameter in the model's reference parameter 
set 
was randomly multiplied by a factor of $1.1$-$1.9$, and the period and waveform 
were examined when the oscillation behavior persists.
The period obtained by increasing the parameters from each reference parameter 
set 
was denoted as $\tau_2$, and the ratio, $\tau_2/\tau_1$, 
was called the relative period. 

\begin{figure}[t]
\begin{center}	
 \hspace*{-0.5cm}
 \includegraphics[width=16cm]{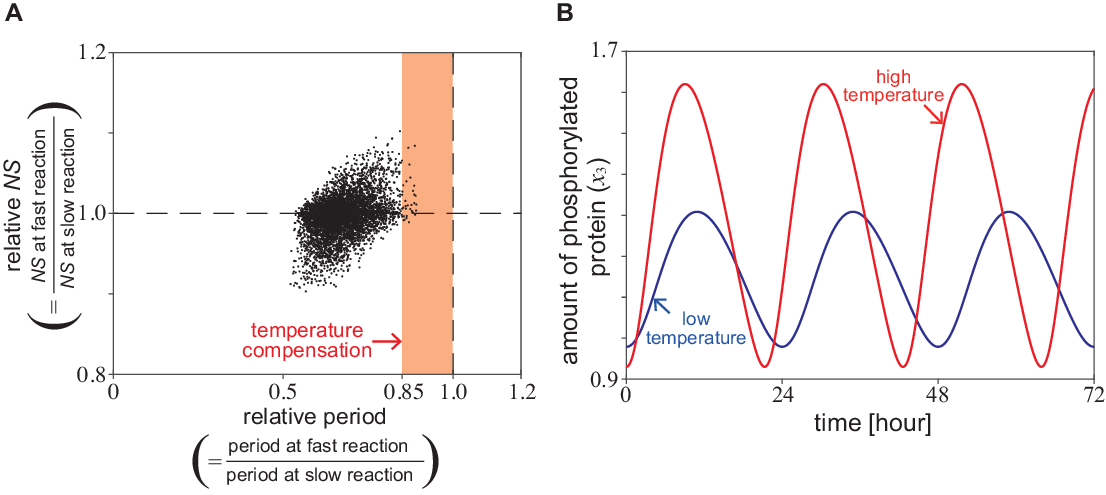}
 \caption{(A) Distribution of relative $NS$ for $x_{3}(t)$ of the circadian 
 clock model 
 as a function of the relative period when reaction rates are increased. 
 We first generated reference parameter sets. To incorporate the effect 
 of the increase of 
 temperature, instead of using the Arrhenius equation, 
 each parameter $k_{1}$, $k_{2}$, $k_{3}$, $p_{1}$, $p_{2}$, 
 and $r$ in the model's reference parameter set was randomly multiplied 
 by a factor of 1.1-1.9. (B) Examples of the waveform in Goodwin model 
 when the period was relatively unchanged. The blue 
 and red lines represent slow and fast reactions, respectively. }	
 \label{figure7}	
 \end{center}
\end{figure}

To quantitatively analyze the correlation between period and waveform 
when temperature compensation occurs, 
we consider
the case of the relative period $\geq 0.85$ because it has been experimentally 
confirmed that the circadian rhythm frequency at high temperature divided by 
that at low temperature of the wild-type ranges 
between $0.85$ and $1.15$ when the temperature
is increased by 10°C and temperature compensation occurs \cite{rub99}. In the 
present numerical analysis, the period 
is relatively stable (relative period $\geq 0.85$) 
in $34$ of the $4900$ parameter sets,
in qualitative agreement with previous theoretical analyses 
that the period often shortens with increasing 
reaction rates \cite{ruoff92, kuro05}. Because the range of parameter 
variation 
is $1.1$-$1.9$ and the average value 
is $1.5$, the reaction rate 
is accelerated by a factor of 1.5 on average, and the average relative period 
is approximately $1/1.5 \approx 0.67$. Figure \ref{figure7} indicates 
that when temperature compensation occurs, there is a clear correlation between the period and waveform.

\subsection{RG analysis of waveform-period correlations}
In the previous section, we demonstrated that the temperature compensation 
of the period 
in the Goodwin model is always accompanied by an increase of the index
$NS$ and waveform distortion correspondingly occurs 
as temperatures increase. 
This raises the following question: 
Is there a universal law governing the waveform distortion occurring when 
temperature increases? To 
answer this question, we derive
an approximate solution for the time evolution of the Goodwin model 
for the circadian rhythm using 
a powerful reduction method, 
called "the renormalization-group(RG) method " \cite{golde89}. 
The solution obtained using the RG method can be interpreted 
as the envelope of the set of solutions given in the perturbation theory, 
which has been applied to various models, including (but not limited to) ODE, PDE, 
discrete systems, and stochastic equations 
\cite{kuni95, ku97, ei20, kuni22, kuni98}. 
To apply the RG method, we again set the transcriptional regulation 
function $f(x_{3})$ to be $r/x_{3}^{n}$. 
In this function, $n$ is the cooperativity of the transcriptional 
regulation, which is a Hopf bifurcation parameter.
The approximate solution of the phosphorylated protein of the circadian clock 
reads (see 
Supplementary Information A.2)
\begin{align}
&x_{3}(t)=\left( \frac{p_{1}p_{2}r}{s_{3}} \right)^{\frac{s_{3}}
{s_{1}s_{2}}}+\varepsilon A_{0}\sin(\omega t)
+\varepsilon^{2} A_{1}A_{0}^{2}\sin(2\omega t+\alpha)+o(\varepsilon^{2})
\label{eqR45}
\end{align}
where we have
\begin{align}
&\varepsilon =n-\frac{s_{4}}{s_{3}},
\label{eqR84} \\
&s_{1}=k_{1}+k_{2}+k_{3}, 
\ s_{2}=k_{1}k_{2}+k_{2}k_{3}+k_{3}k_{1}, 
\label{eqR50} \\
&s_{3}=k_{1}k_{2}k_{3}, 
\  s_{4}=(k_{1}+k_{2})(k_{2}+k_{3})(k_{3}+k_{1}),
\label{eqR51}
\end{align}
with the angular velocity and the phase parameter of the second-order term
\begin{align}
&\omega = \sqrt{s_{2}}-\varepsilon\frac{s_{1}s_{3}s_{4}}{6(2s_{1}s_{2}^{2}-
(s_{1}^{2}+6s_{2})s_{3})\sqrt{s_{2}}}+o(\varepsilon^2),
\label{eqR48} \\
&\alpha = \arctan\left( \frac{s_{1}}{2\sqrt{s_{2}}} \right),
\label{eqR49}
\end{align}
as well as the amplitudes
\begin{align}
&A_{0}=\sqrt{\frac{4((s_{1}^{2}+s_{2})^{2}-2\varepsilon s_{1}s_{3})
(s_{1}^{2}+4s_{2})s_{3}^{3}}
{\varepsilon (2s_{1}s_{2}^{2}-(s_{1}^{2}+6s_{2})s_{3})
(s_{1}^{2}+s_{2})^{2}s_{4}s_{1}s_{2}}}
\left( \frac{p_{1}p_{2}r}{s_{3}} \right) ^{\frac{s_{3}}{s_{1}s_{2}}},
\label{eqR46} \\
&A_{1}=\frac{s_{4}s_{1}}{12s_{3}\sqrt{s_{1}^2+4s_{2}}}
\left( 
\frac{s_{3}}{p_{1}p_{2}r}
\right) ^{\frac{s_{3}}{s_{1}s_{2}}}.
\label{eqR47}
\end{align}

\begin{figure}[t]	
 \begin{center}	
 \hspace*{-1cm}
  \includegraphics[width=16cm]{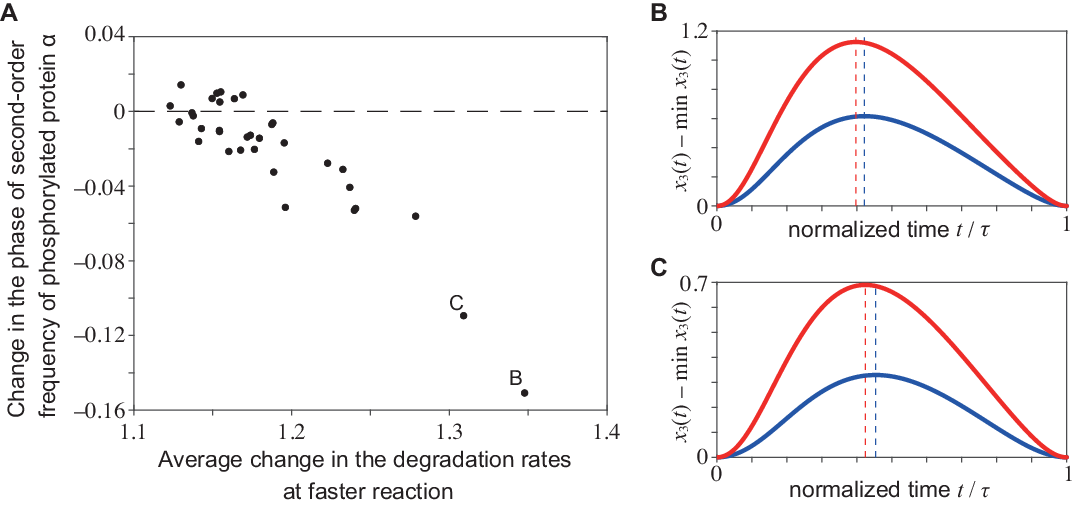}	
  \caption{Change in the phase of the second-order frequency $\alpha$ in 
  phosphorylated protein oscillation under varying circadian clock 
  model parameters. 
  (A) The parameters $k_{1}$, $k_{2}$, $k_{3}$, $p_{1}$, $p_{2}$, and $r$ 
  were randomly increased from $1.1$ to $1.9$. The horizontal axis represents 
  the arithmetic mean of the fast-case values of $k_{1}$, $k_{2}$, and $k_{3}$ 
  divided by the slow-case values. 
  The parameters $p_{1}$, $p_{2}$, and $r$ 
  are omitted because they do not affect the period in Eq. \eqref{eqR48}. 
  The vertical axis presents changes in the phase of the second-order 
  frequency obtained via generalized harmonic analysis of numerically calculated 
  periodic time series. When the period length is maintained 
  (relative period within $0.85$-$1.0$), the phase of $\alpha$ of $x_{3}(t)$ tends 
  to decrease as the reaction rate increases. (B, C) Time series 
  for the increased reaction rates ($1.1$-$1.9$) while maintaining the period. 
  The red and blue lines 
  represent the time series for increased reaction rates 
  and 
  those for the reference parameter set, respectively. 
  The parameter values are provided in Supplementary Table 3. 
  For comparison, the period length is standardized to $1$. 
  The time of the minimum value is set to $0$, and the maximum phosphorylated 
  protein oscillation is indicated by the dotted line. 
  Increased reaction rates tend to advance the timing of the maximum oscillation.}
   \label{figure4}	
 \end{center}
\end{figure}

The RG method provides an approximate but globally valid solution, 
and thus enable us to make a 
detailed investigation of 
the waveform distortion when temperature compensation occurs 
in the Goodwin model. 
The numerical analysis using the same parameter sets as used in Fig. \ref{figure7}A, 
in which the relative period in the model remains stable
and within the interval $(0.85,\,1.0)$ against the temperature variations, 
shows that the phase parameter $\alpha$ in the 2nd-order frequency
tends to decrease with increasing reaction rates
(see Fig. \ref{figure4}A). 
When the increase in reaction rates is small, the change 
in the phase of the second-order frequency scatters around zero 
and is negligible.
However, with a significant increase in the reaction rates, 
the phase of the second order always tends 
to decrease as the reaction rates increase.

The significance of the phase parameter $\alpha$ given in the second-order term 
on the waveform of the time series can be understood intuitively as follows: 
when the phase $\alpha$ is large, the increasing duration tends to become 
longer because of
the less overlap of the time profiles given by $\sin(\omega t)$ and 
$\sin(2\omega t+\alpha)$ (Fig. \ref{figure4}BC, blue line). 
Conversely, a smaller $\alpha$ tends to result 
in a shorter increasing duration because of an additive effect of the two terms, 
which leads to a steeper slope on total, as presented 
in Fig. \ref{figure4}BC, red line. 
Therefore, the numerical results in Fig. \ref{figure4}A suggest 
that the decreasing duration 
of the time series elongates with as the reaction rate 
increases when the period is relatively stable despite 
the increasing reaction rate.

\begin{figure}[t]	
 \begin{center}
  \includegraphics[width=16cm]{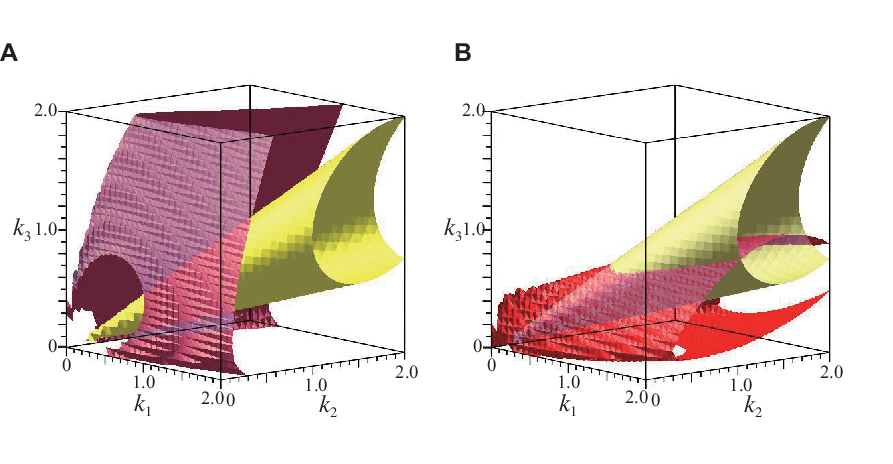}	
  \caption{The parameter space for the constant period ($d\omega =0$, 
  \textit{red surface}) with increasing degradation rates $k_{1}$, $k_{2}$, 
  and $k_{3}$,
  and that for the constant phase of second-order frequency ($d\alpha =0$, 
  \textit{yellow surface}). The phase of second-order frequency $\alpha$ should 
  decrease with increasing $k_{1}$, $k_{2}$, and $k_{3}$ outside the yellow 
  surface ($d\alpha <0$). The cooperativity $n$ is $n=12$ (A), and $n=20$ (B), 
  and the variations of the parameters are set to $dk_{1}=0.3$, $dk_{2}=0.2$, 
  and $dk_{3}=0.1$. (A) When $n$ is small ($n=12$), surface $d\omega =0$ is 
  included in space $d\alpha <0$, which means that $\alpha$ should become smaller 
  if $\omega$ does not change with increasing $k_{1}$, $k_{2}$, 
  and $k_{3}$. (B) When $n$ is large ($n=20$), 
  surface $d\omega =0$ can intersect $d\alpha =0$ 
  although high cooperativity is unrealistic.}
 \label{figure5}
 \end{center}
\end{figure}

For a theoretical confirmation of  the numerical result that the 
phase parameter $\alpha$ given in the second-order term
tends to become smaller as the temperature increases 
when the period is temperature-compensated, 
we analyze
the sensitivity of the angular frequency $\omega$ and the phase $\alpha$ 
to the reaction rates by utilizing
the results of the RG method.
With use of Eqs. \eqref{eqR48} and \eqref{eqR49}, we calculate
$d\omega =(\partial\omega/\partial k_{1})dk_{1}+(\partial\omega/\partial 
k_{2})dk_{2}+(\partial\omega/\partial k_{3})dk_{3}$ and $d\alpha =
(\partial\alpha/\partial k_{1})dk_{1}+(\partial\alpha/\partial k_{2})dk_{2}+
(\partial\alpha/\partial k_{3})dk_{3}$. 
In Fig. \ref{figure5}A, we present the parameter regions given 
by the constraints $d\omega =0$ (red surface) and 
$d\alpha =0$ (yellow surface) for 
the cooperativity $n=12$ of the transcription regulation. 
We can see that the region for $d\alpha<0$ (outside yellow surface) 
includes that given 
by the constraint $d\omega =0$ for all $k_{i}$ ($i=1,\,2,\,3$).
This implies that if the period is robust against a change 
in the parameters 
$k_{i}$ ($i=1,\,2,\,3$), then the phase $\alpha$ in the second-order term 
always becomes smaller with increasing parameters.

Next, let us examine how the parameter regions given by the constraints $d\omega =0$ and $d\alpha =0$ 
change with variations of the cooperativity $n$. 
The numerical calculation 
shows that the region corresponding to $d\alpha <0$ 
includes that given 
by $d\omega =0$ for $n=13$ and $14$. 
However, in the case of an exceedingly high cooperativity of 
transcription regulation $n$, which is not biologically realistic, 
$d\alpha$ can be positive when $d\omega =0$. For instance, for $n=20$, 
although $d\alpha$ is negative for most of the parameter space, 
there {\em is} a region in which $d\alpha >0$ when $d\omega =0$ 
(see Fig. \ref{figure5}B).
These results indicate that if circadian rhythms are stable 
under temperature variations, 
the slope in the increasing phase of phosphoproteins should become sharper as 
temperature increases.
Thus, we conclude that 
(i) the waveform of the gene activity rhythm should be more distorted 
at higher temperatures, and (ii) the rate of the increase 
in phosphoprotein levels 
should be greater at higher temperatures if temperature compensation is achieved. 
In principle, these features can be tested experimentally.

\subsection{Verification of the theoretical analysis of temperature 
compensation using published experimental data}
The period formula Eq. \eqref{eq:tau} of the Goodwin model
indicates that the non-sinusoidal index $NS$ of the waveform 
of the circadian rhythms becomes larger, 
implying greater distortion of the waveform when all reactions are faster at 
higher temperatures during temperature compensation. 
To test this theoretical prediction of circadian gene activity in actual 
organisms, we analyze
the waveform of the activity rhythms of the timeless gene in Drosophila at $18$ and 
$29$°C using published experimental data \cite{kidd15}.

\begin{figure}
 \begin{center}
  \includegraphics[width=15cm]{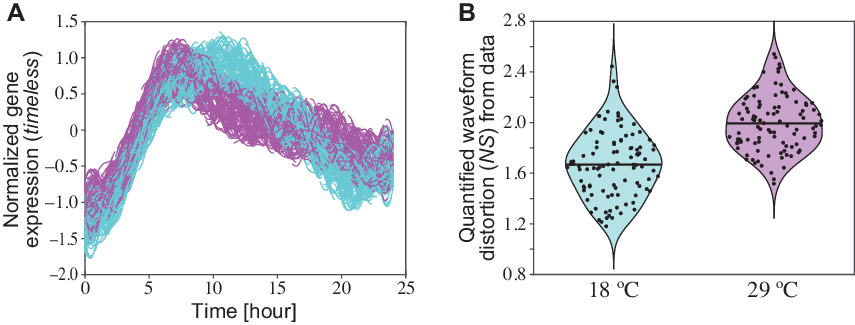}	
   \caption{Analysis of the circadian waveform of wild-type 
   Drosophila at different temperatures using previously reported 
   experimental data \citep{kidd15}. (A) Re-plot of Fig. 3C in \citep{kidd15}. 
   The average curves of \textit{tim-luc} at $18$ and $29$ ${}^\circ$C 
   are extracted, and 
   uniformly distributed noise 
   between $-0.4$ and $0.4$ 
   is added to the extracted data, 
   generating $100$ time series datasets. 
   Spline interpolation 
   is applied 
   to set the sampling interval to $0.1$ h. 
   The interpolated time series 
   data at $18$ (cyan) and $29$ ${}^\circ$C (magenta) are plotted. (B) Distribution 
   of the quantified waveform distortion ($NS$) from the data 
   at $18$ (cyan) and $29$ ${}^\circ$C (magenta). The time series data with noise 
   are detrended by 
   multiplying 
   an exponential function so that the positions of local minima of the oscillations 
   are approximately equal. 
   The Fourier coefficients of the detrended time series 
   are quantified using GHA. The $NS$ values 
   are estimated from the coefficients
   up to the third harmonics. Dots represent the $NS$ values for each dataset, 
   and the horizontal lines represent their average values.}
 \label{figure6}	
 \end{center}
\end{figure}

First, we extract the time series of the average curve 
from Fig. 3C in a prior study \citep{kidd15} 
using WebPlotDigitizer per hour. 
Second, we add uniformly distributed noise between 
$-0.4$ and $0.4$ 
to the extracted 
data to consider data errors. Then, we interpolate
the time series every $0.1$ h 
using spline interpolation (Fig. \ref{figure6}A).
The interpolated data were detrended by multiplying 
an exponential 
function so that the position of the local minima of the oscillations 
are approximately reproduced. 
Then, the detrended time series is fitted with 
a sum of trigonometric functions up to the third harmonics using 
the generalized harmonic analysis (GHA) method \cite{kuro19, wiener58, terada94, gibo20}. 
The width of the window for the analysis is set to one period.
Using the Fourier coefficients of the fitting time series, 
we 
evaluate the distribution and average value of $NS$. 
The 
resultant $NS$ of the activity rhythms 
of the timeless gene, as defined by Eq. \eqref{eq1}, 
at a higher temperature ($29$ ${}^\circ$C) tend to be larger than 
that at a lower temperature ($18$ ${}^\circ$C), whereas
the $NS$ values are somewhat varied (Fig. \ref{figure6}B), 
which 
is consistent with 
the prediction.
Experimental studies have demonstrated that temperature compensation 
can be impaired by genetic mutations. In the Drosophila mutant perL, 
the period increases with increasing temperature \cite{kidd15, rosbash95}. 
Equation \eqref{eq:tau} implies that if the period increases with temperature, 
then the waveform of the circadian rhythm should be non-sinusoidal and more 
distorted at higher temperatures. 
Thus, it is predicted that the waveform 
of the circadian gene activity in perL should become more non-sinusoidal 
with higher temperatures. Again, we can quantify the waveform of perL using 
experimental data \cite{kidd15} (Supplementary Fig. \ref{figure1}). The waveform 
of circadian gene activity tends to be more non-sinusoidal 
at higher temperatures in perL, as observed in the wild-type, 
whereas the $NS$ values varied, in line with the prediction.

\section{Theoretical analysis of synchronization in the circadian rhythm model}
\begin{figure}[t]
 \begin{center}	
 \hspace*{0cm}
  \includegraphics[width=16cm]{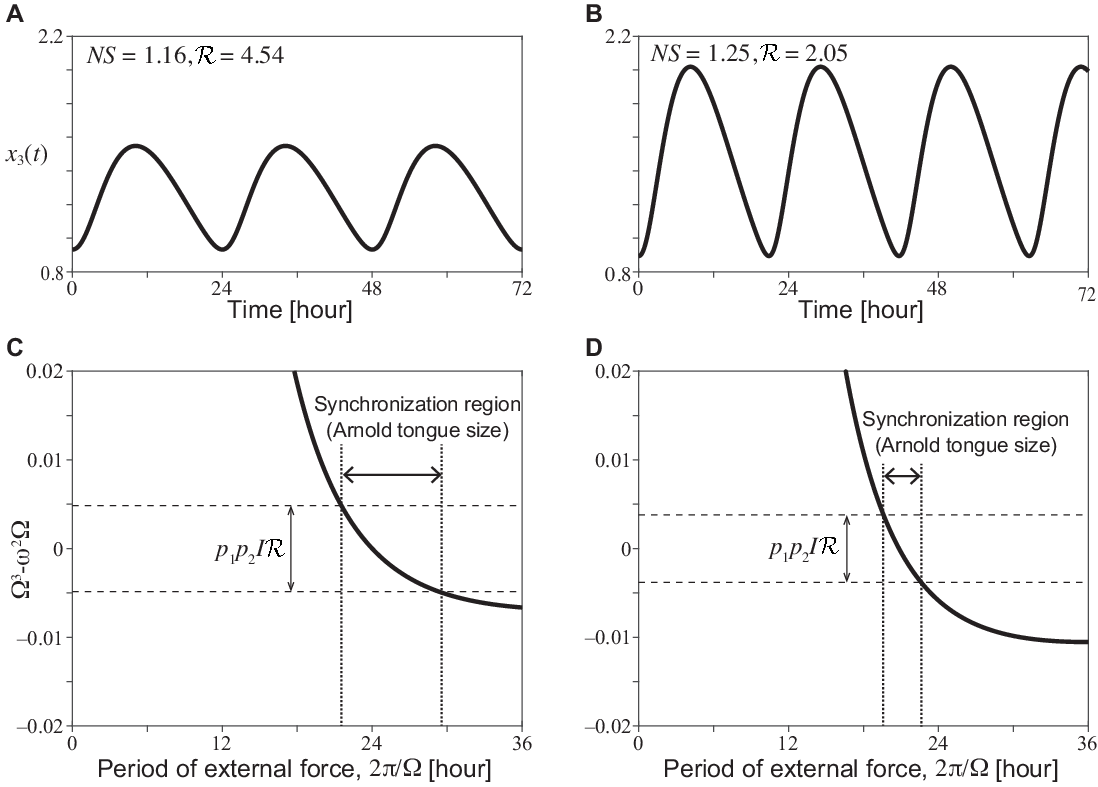}	
  \caption{The time series depicts cases in which the period length remains relatively constant despite increasing reaction rates in the circadian clock model (A: reference set case, B: increased reaction rates case). Additionally, synchronization regions for these cases are presented (C: reference set case, D: increased reaction rates case). Panels A and B correspond to Fig. \ref{figure4}B, but they include additional waveform values such as $NS$ and ${\cal R}$. Synchronization regions for $I=0.03$ in Eq. \eqref{eq30} (C, D), calculated using Eq. \eqref{eq35}, are illustrated for each oscillation (A and B). Synchronization with the external forcing period occurs only when the cubic function $\Omega^{3} - \omega^{2}\Omega$ falls within a specific range. This range becomes narrower as the waveform distortion increases along with reaction rates, consistent with the observations in Fig. \ref{figure3}B.}	
 \label{figure2}
 \end{center}
\end{figure}

The numerical 
reslut in Fig. \ref{figure3}B 
shows that the range of synchronization into the light-dark cycles tends 
to decrease as the waveform becomes more distorted 
in the simple circadian clock model.
To 
clarify the condition for synchronization, we again 
consider
the Goodwin model but with an external 
force incorporated as follows:

\begin{align}
&\frac{dx_{1}}{dt}=f(x_{3})-k_{1}x_{1}+I\cos(\Omega t)
\label{eq30} \\
&\frac{dx_{2}}{dt}=p_{1}x_{1}-k_{2}x_{2}
\label{eq31} \\
&\frac{dx_{3}}{dt}=p_{2}x_{2}-k_{3}x_{3}
\label{eq32}
\end{align}
where $I\cos(\Omega t)$ is a periodic environmental change, 
such as a light-dark cycle. By eliminating $x_{1}$ and $x_{2}$, 
Eqs. \eqref{eq30}-\eqref{eq32} is converted to the following single equation:
\begin{equation}
\frac{d^{3}x_{3}}{dt^{3}}+s_{1}\frac{d^{2}x_{3}}{dt^{2}}
+s_{2}\frac{dx_{3}}{dt}+s_{3}x_{3}
=p_{1}p_{2}f(x)+p_{1}p_{2}I\cos(\Omega t).
\label{eq33}
\end{equation}
If the model is to admit a synchronization 
to the external cycle $I\cos(\Omega t)$ at all, 
then $x_{3}(t)$ should be written as the Fourier series 
$x_{3}(t)=\sum_{j=-\infty}^{\infty}a_{j}\exp(i\Omega jt)$.
Multiplying Eq. \eqref{eq33} by $dx/dt$ and integrating 
that for the interval $t$ to $t+2\pi/\Omega$, we have the following equation:
\begin{equation}
\Omega^{3}-\omega^{2}\Omega = 
\frac{1}{2}p_{1}p_{2}I{\cal R}\sin \beta
\label{eq34}
\end{equation}
where 
\[
\omega=2\pi/\tau=\sqrt{s_{2}\sum_{j=1}^{\infty}|a_{j}|^{2}j^{2}/\sum_{j=1}^{\infty}|a_{j}|^{2}j^{4}}
\]
is the natural angular frequency without the external force and 
\begin{equation}
{\cal R}=
\frac{|a_{1}|}{\sum_{j=1}^{\infty}|a_{j}|^{2}j^{4}}
\label{eq:calA}
\end{equation}
 with $\beta$ being the argument of $a_{1}$ such that $a_{1}=|a_{1}|\exp(i\beta)$. 
Because $-1\leq \sin\beta\leq 1$, when $x(t)$ synchronizes with the external 
cycles, the angular frequency of the external cycles ($\Omega$) should satisfy
the inequality

\begin{equation}
|\Omega^{3}-\omega^{2}\Omega |
\leq \frac{1}{2}p_{1}p_{2}I{\cal R}.
\label{eq35}
\end{equation}
We note that
${\cal R}$ defined by Eq. \eqref{eq:calA} becomes smaller when the components 
of higher harmonics become larger and the waveform exhibits greater
distortion. Therefore, if the waveform is more distorted 
by, say, an increasing temperature, the bounds of Eq. \eqref{eq35} become smaller,
and accordingly, the allowed region of the middle term is narrower. 
The left hand side of Eq. \eqref{eq35} is a cubic function of $\Omega$, 
which is monotonically increasing near $\Omega=\omega$. 
If the waveform is more distorted, 
 Eq.(\ref{eq:calA})
should be smaller, making the allowed region of Eq. \eqref{eq35} narrower. 
Then, the range of $\Omega$ that causes synchronization becomes narrower, 
as presented in Fig. \ref{figure2}. 
This indicates that the range of synchronization into light-dark cycles always
decreases as the waveform becomes more distorted in the simple circadian model, 
which is consistent with the numerical simulation in Fig. \ref{figure3}B.

\section{Discussion}

We theoretically explored the conditions for clarifying the temperature 
compensation of the biological clock and its synchronization to light-dark cycles 
with a particular focus on waveform distortion. The theoretical analysis 
of the Goodwin model, one of the most studied models of biological 
clocks, revealed that waveform distortion of gene activity rhythms with increasing 
temperature is necessary for temperature compensation. Furthermore, we derived an 
approximate but globally valid solution to the waveform of the time profiles using 
the RG method as a powerful tool for global analysis. This allowed us to 
investigate, based on the analytical solution, whether there is a universal law 
for the mechanism by which the waveform changes with temperature 
variation.

The results indicated that temperature compensation is more likely to 
occur if the waveform is distorted if the decreasing duration of 
circadian protein oscillation elongates as temperature increases. 
Although theoretical predictions based on a model might not 
always be realized in real organisms, we quantified the gene activity rhythms of 
published experimental data using Drosophila. This quantification confirmed that 
the waveform is distorted at high temperatures, in accordance with our 
theoretical predictions.

It is notable that the systematic wave distortion governed by Eq. (\ref{eq:tau}), 
which we have found to hold in the Goodwin model, also applies to a wide 
class of non-linear oscillators used for biological phenomena different from 
biological rhythms, including the Lotka-Volterra model \cite{murray88}, 
which is commonly used in ecology, and the van der Pol model, as 
presented in the Supplementary Infomration A.3 and A.4 (see also \cite{kuro19}). This suggests that 
exploring the possible significance of waveform distortion in other mathematical 
models, such as the Fitz-Hugh-Nagumo model in neuroscience, would be intriguing 
\cite{murray88, fitz61, nagumo62}.

To the best of our knowledge, this is the first study to apply the RG method, 
a powerful resummation method of the perturbation series first developed in physics, 
to circadian rhythm problems. In the RG method, secular terms appearing in the 
na\"{i}ve perturbation series are renormalized into the 'integral constants' 
, which thus acquire the nature of the slow modes, 
making it a powerful tool for global and asymptotic analysis. 
Unlike the naive perturbation theory, the solutions given by the RG method provide 
a time evolution close to numerical simulations in 
the relevant global domain of time, 
offering an approximate solution for the period and waveform. 
The analytical results predict that longer tails of gene activity rhythms 
at higher temperatures occur for temperature compensation.

This study
also investigated the synchronization 
with environmental light-dark cycles at various temperatures \cite{pikovsky03, abra10}. The numerical simulations and theoretical analysis predict 
that as the distortion 
of the gene activity rhythms for achieving a temperature-compensated period 
increases, it becomes more difficult to synchronize with 
the light-dark cycle. 
This prediction aligns with the reported temperature-dependent variation 
in response 
to light pulses in Drosophila and Neurospora, 
displaying smaller phase shifts 
at higher temperatures \citep{nak80} \citep{varma13}.

As mentioned in the Introduction, previous theoretical and experimental 
studies, such as those using Drosophila \cite{kidd15}, suggested 
that waveforms in gene activity rhythms do not change under temperature variations, 
although they are blurred by error bands. These findings are 
apparently inconsistent with our current conclusion. We believe that this 
discrepancy stems from two main factors: differing assumptions about the 
temperature sensitivity of degradation rates and differing interpretations of 
experimental results regarding gene activity rhythms at distinct temperatures. 
First, the previous study assumed that all degradation rates are 
temperature-insensitive. Thus, the waveform of gene activity rhythms does not need 
to change with temperature. By contrast, we assume that some degradation 
rates at least should accelerate with temperature, leading to a more distorted 
waveform of gene activity rhythms at higher temperatures. Second, the previous 
study \cite{kidd15} interpreted their experimental results as indicating that gene 
activity rhythms at different temperatures can be collapsed onto each other by 
rescaling, supporting their prediction that temperature compensation occurs because 
of rescaling and the stability of the temperature insensitivity of degradation 
rates. Conversely, our quantification of their experimental data 
indicated that the waveform tends to be more distorted at higher 
temperatures, whereas variation in $NS$ values was noted. Thus, the 
present result is consistent 
with our theoretical 
prediction. We believe that further systematic quantification of the waveforms of 
gene activity and/or protein activity rhythms in various circadian organisms will 
be essential for clarifying the importance of the waveform in circadian 
rhythms in the future.

\section*{Methods}
\subsection*{Computation of ODEs}
The ordinary differential equations in this work were calculated by a fourth-order Runge-Kutta method with MATLAB (The MathWorks, Natick, MA). The time step size $\Delta t$ for the Goodwin model (Eqs. \eqref{eq:G1}-\eqref{eq:G3}) was $0.001$, and that for the Lotka-Volterra model (Eqs. \eqref{eq2}-\eqref{eq3}) and the van der Pol oscillator (Eq. \eqref{eqR92}) was $0.0001$.

\subsection*{Numerical analysis of synchronization with light-dark cycles}
The angular frequency of the light-dark cycle $\Omega$ in Eq. \eqref{eq:G1-forced} was changed by $0.0001$. We considered that the model was synchronized with the light-dark cycle when the change in the amplitude was smaller than the rounding error as in Fig. 1B.

\subsection*{Estimation of waveform distortion from simulated time-series}
We estimated waveform distortion, i.e., non-sinusoidal power ($NS$, see Eq. \eqref{eq1}) from the simulated time-series by using the generalized harmonic analysis (GHA) method as in Figs. 1B and 2A. The GHA method estimates the amplitudes $A_{j}$ and $B_{j}$, and the frequencies $f_{j}$, which minimize the squared residual $\int_{0}^{L}[x(t)-\sum_{j=1}^{j_{max}}\{ A_{j}\cos(2\pi f_jt)+B_{j}\sin(2\pi f_{j}t) \} ]^{2}dt$. The detailed procedure of GHA can be found in \cite{kuro19, terada94}.

\subsection*{Estimation of phases of 
Fourier components from simulated time-series}
By using GHA method, we 
computed the Fourier series $x(t)=\sum_{j=1}^{j_{max}}\{ A_{j}\cos(2\pi f_jt)+B_{j}\sin(2\pi f_{j}t)\}$. This form was converted into $x(t)=\sum_{j=1}^{j_{max}}r_{j}\sin(2\pi f_{j}t+\alpha_{j})$, where $r_{j}=\sqrt{A_{j}^{2}+B_{j}^{2}}$ and $\alpha_{j} =\arctan (B/A)$.

\subsection*{Computation of parameter space for temperature compensation using the RG method}
The frequency change $d\omega =(\partial\omega/\partial k_{1})dk_{1}+(\partial\omega/\partial k_{2})dk_{2}+(\partial\omega/\partial k_{3})dk_{3}$ and the phase change $d\alpha =(\partial\alpha/\partial k_{1})dk_{1}+(\partial\alpha/\partial k_{2})dk_{2}+(\partial\alpha/\partial k_{3})dk_{3}$ were calculated from Eqs. \eqref{eqR48} and \eqref{eqR49} by using Maple. The parameter space for $d\omega =0$ and $d\alpha =0$ in Fig. \ref{figure5}AB were numerically generated by Maple.

\subsection*{Estimation of waveform distortion from published experimental data}
The average curves of experimental time-series of \textit{timeless} luciferase (\textit{tim-luc}) reporter (shown to recapitulate the dynamics of \textit{timeless} gene expression) at 18 and 29 ${}^\circ$C of Figs. 3C and 4B in the published literature \cite{kidd15} were extracted using WebPlotDigitizer at $1$ h intervals. Uniformly distributed noise between $-0.4$ and $0.4$ was added to the extracted data, generating $100$ time series datasets so that the experimental noise 
is roughly reproduced as in Fig. 5A and Supplementary Fig. 3A. Spline interpolation was applied to set the sampling interval to $0.1$ h to smooth time series data. The time series data with noise were detrended by multiplying an exponential function so that the positions of local minima of the oscillations are approximately
reproduced. The Fourier coefficients of the detrended time series were quantified using GHA. The $NS$ values were estimated from the coefficients up to the third harmonics as in Fig. 5B and Supplementary Fig. 3B.

\section*{Acknowledgements}
We thank H. Nakao, H. Chiba, Y. Kawahara, A. Mochizuki for useful comments on this study. This work was supported by grants from Japan Science and Technology Agency (JPMJCR1913 to G.K.), and from the Japanese Society for the Promotion of Science, and the Ministry of Education, Culture, Sports, Science, and Technology in Japan (JP21K06105 to G.K., 19K03872 to T.K.).

\bibliography{getwriting}

\newpage
\setcounter{page}{1}
\section*{Supporting Information for}

\section*{Waveform distortion for temperature compensation and synchronization in circadian rhythms: An approach based on the renormalization group method}

Shingo Gibo$^{1*}$, Teiji Kunihiro$^{2}$, Tetsuo Hatsuda$^{1}$, and Gen Kurosawa$^{1*}$\\

$^{1}$Interdisciplinary Theoretical and Mathematical Sciences Program (iTHEMS), RIKEN \\
$^{2}$Yukawa Institute for Theoretical Physics (YITP), Kyoto University

$^{*}$Correspondence: shingo.gibo@riken.jp (S.G), g.kurosawa@riken.jp (G.K)

\textbf{This PDF includes:} \\
Supplementary Text \\
Supplementary Figures 1 to 3 \\
Supplementary Tables 1 to 2

\newpage
\renewcommand{\figurename}{Supplementary Figure}
\setcounter{figure}{0}
\section*{Supplementary Text}
\subsection*{A.1 Brief introduction of the renormalization group (RG) method 
using a simple model with Hopf bifurcation}
In this section, we introduce the RG method \cite{golde89} in a geometrical manner as 
formulated in \cite{kuni95, ei20} with a simpler prescription 
without the redundant 'time-splitting' procedure.
In this aim, we 
use a generic model with a Hopf bifurcation. 
A more detailed account of
the method is given in \cite{kuni22}.

Let us consider the model equation 
\begin{equation}
Lx(t)=F(x(t); \varepsilon)
\label{eqR1}
\end{equation}
where $x(t)$ is a state variable of our dynamics, 
$
L=\sum_{n=1}^{N}a_{n}(d/dt)^{n}
$
 is a linear differential operator, $F(x(t); \varepsilon)$ is a nonlinear function of $x(t)$, 
 and $\varepsilon$ is an internal parameter of $F$, which acts as a bifurcation parameter of the system. 
We assume that the model has a fixed point $x_{0}$ satisfying the equation
$F(x_{0}; 0)=0$, which is destabilized for $\varepsilon >0$ through 
the Hopf bifurcation. We are interested in the derivation of the reduced equation 
and an approximate but valid solution in a global domain of time around the critical point 
of the Hopf bifurcation.
Thus,
we apply
the perturbation theory and express the solution around an arbitrary 
time $t=t_{0}$ belonging to a global domain in the asymptotic regime (see below) in a power series of $\varepsilon$ as follows:
\begin{equation}
x(t;t_{0})=x_{0}+\varepsilon u_{1}(t;t_{0})+\varepsilon ^{2}u_{2}(t;t_{0})+\varepsilon^{3}u_{3}(t;t_{0})+o(\varepsilon^{3}).
\label{eqR2}
\end{equation}
Substituting Eq. \eqref{eqR2} into Eq. \eqref{eqR1} and equating the terms with the same powers of $\varepsilon$, we obtained
\begin{align}
&O(\varepsilon^{1}): L'u_{1}=0,
\label{eqR3} \\
&O(\varepsilon^{2}): L'u_{2}=f_{1}(u_{1}),
\label{eqR4} \\
&O(\varepsilon^{3}): L'u_{3}=f_{2}(u_{1},u_{2}),
\label{eqR5}
\end{align}
where 
\[
L'=L-(\partial F/\partial x)|_{x=x_{0},\varepsilon =0},
\]
and $f_{1}(u_{1})$ and $f_{2}(u_{1},u_{2})$ are nonlinear functions that depend on $F(x_0)$. 
Because Hopf bifurcation occurs at $\varepsilon =0$, two of the eigenvalues of the linear differential operator $L'$ are written as 
$\pm i\omega_{0}$, with $\omega_0$ being a real number, and the others have negative real parts as
$Re(\lambda_{k})<0$ ($k=1,...,N-2$). 

Then, the first-order solution can be expressed as
\begin{equation}
u_{1}(t;t_{0})=A(t_{0})\cos(\omega_{0} t+\theta (t_{0}))+\sum_{k=1}^{N-2}c_{k}(t_{0})e^{\lambda_{k}t},
\label{eqR6}
\end{equation}
where $A(t_{0})$, $\theta (t_{0})$, and $c_{k}(t_{0})$ are the 
integral constants which are assumed to
depend on the initial time $t_{0}$. 

Next, we consider the asymptotic regime as $t\,\to\, \infty$ so that 
the second term describing the transient behavior has virtually become negligible. 
Then,
the first-order solution in this asymptotic regime can be expressed only by the first term as
\begin{equation}
u_{1}(t \rightarrow \infty ;t_{0})=A(t_{0})\cos(\omega_{0} t+\theta (t_{0})).
\label{eqR7}
\end{equation}

Next, we proceed to the second-order equation. 
Substituting Eq. \eqref{eqR7} into Eq. \eqref{eqR4}, we have
\begin{equation}
L'u_{2}
=b_{1}A\cos(\omega_{0}t+\theta)
+b_{2}A^{2}\cos(2(\omega_{0}t+\theta))
+b_{2}A^{2},
\label{eqR8}
\end{equation}
where $b_{k}$ ($k=1, 2$) are constants depending on $f_{1}(u_{1})$. 
It is to be noted that the inhomogeneous part (r.h.s.) contains a term proportional to $\cos(\omega_{0} t+\theta)$, which is a zero mode of the linear operator that gives rise to secular terms in the particular solutions of the inhomogeneous equation.
The general solution to Eq. \eqref{eqR8} is given as a sum of a particular solution to the inhomogeneous equation 
and the general solution to the homogeneous equation.
Now, it is possible and convenient to choose 
the coefficients of the latter so that all of the secular terms vanish
 at $t=t_0$ \cite{kuni95}, which leads to the second-order solution as

\begin{align}
u_{2}(t;t_{0})=&(t-t_{0})d_{1}A\cos(\omega_{0} t+\theta)
+(t-t_{0})d_{2}A\sin(\omega_{0} t+\theta) \notag \\
&+d_{3}A^{2}\cos(2(\omega_{0}t+\theta))
+d_{4}A^{2}\sin(2(\omega_{0}t+\theta))
+d_{5}A^{2}, 
\label{eqR9}
\end{align}
where $d_{k}$ ($k=1\cdots 5$) are constants depending on the right-hand side of Eq. \eqref{eqR8}. 

Similarly, the third-order solution takes the form of
\begin{align}
u_{3}(t;t_{0})=&(t-t_{0})(f_{1a}A^{3}+f_{1b}A)\cos(\omega_{0}t+\theta)
+(t-t_{0})(f_{2a}A^3+f_{2b}A)\sin(\omega_{0}t+\theta) \notag \\
&+(t-t_{0})^{2}f_{3}A\cos(\omega_{0}t+\theta)
+(t-t_{0})^{2}f_{4}A\sin(\omega_{0}t+\theta)
+f_{5}A^{2}\cos(2(\omega_{0}t+\theta)) \notag \\
&+f_{6}A^{2}\sin(2(\omega_{0}t+\theta))
+(t-t_{0})f_{7}A^{2}\cos(2(\omega_{0}t+\theta))
+(t-t_{0})f_{8}A^{2}\sin(2(\omega_{0}t+\theta)) \notag \\
&+f_{9}A^{3}\cos(3(\omega_{0}t+\theta))+f_{10}A^{3}\sin(3(\omega_{0}t+\theta))
+f_{11}A^{2}+(t-t_{0})f_{12}A^{2},
\label{eqR87}
\end{align}
where $f_{1a}$, $f_{1b}$, $f_{2a}$, $f_{2b}$, and $f_{k}$ ($k=3\cdots 12$) are constants depending on Eqs. \eqref{eqR5}, \eqref{eqR7}, and \eqref{eqR9}.
Note that the solution is constructed so that the secular terms 
vanish at $t=t_0$. 

Thus, collecting all of the terms, the approximate solution to Eq. \eqref{eqR1} up to the third order 
of $\varepsilon$ reads
\begin{align}
x(t;t_{0})=&x_0+\varepsilon A\cos(\omega_{0} t+\theta)
+\varepsilon^{2}\{(t-t_{0})d_{1}A\cos(\omega_{0} t+\theta)
+(t-t_{0})d_{2}A\sin(\omega_{0} t+\theta) \notag \\
&
+d_{3}A^{2}\cos(2(\omega_{0}t+\theta))
+d_{4}A^{2}\sin(2(\omega_{0}t+\theta))
+d_{5}A^{2}\} \notag \\
&+\varepsilon^{3}\{ (t-t_{0})(f_{1a}A^{3}+f_{1b}A)\cos(\omega_{0}t+\theta)
+(t-t_{0})(f_{2a}A^3+f_{2b}A)\sin(\omega_{0}t+\theta) \notag \\
&+(t-t_{0})^{2}f_{3}A\cos(\omega_{0}t+\theta)
+(t-t_{0})^{2}f_{4}A\sin(\omega_{0}t+\theta)
+f_{5}A^{2}\cos(2(\omega_{0}t+\theta)) \notag \\
&+f_{6}A^{2}\sin(2(\omega_{0}t+\theta))
+(t-t_{0})f_{7}A^{2}\cos(2(\omega_{0}t+\theta))
+(t-t_{0})f_{8}A^{2}\sin(2(\omega_{0}t+\theta)) \notag \\
&+f_{9}A^{3}\cos(3(\omega_{0}t+\theta))+f_{10}A^{3}\sin(3(\omega_{0}t+\theta))
+f_{11}A^{2}+(t-t_{0})f_{12}A^{2}
\} +o(\varepsilon^{3}).
\label{eqR10}
\end{align}
Because Eq. \eqref{eqR10} contains the secular terms, 
this solution 
is valid only locally around $t=t_{0}$,
but 
it exhibits a divergent behavior as $\vert t-t_0\vert$ 
goes infinity. 
In fact, this is a rather common behavior occurring in na\"{i}ve perturbation expansions.

Next, we use a geometrical viewpoint to circumvent the disastrous situation 
following \cite{kuni95}.
The solution \eqref{eqR10} gives a family of curves with $t_0$ being the parameter specifying each curve 
in the $t$-$x$ plane. Each curve gives a good approximate solution to the original equation in a local 
domain around $t=t_0$. The idea is that the {\em envelope curve} of the family of curves
hopefully gives an approximate but valid solution in the global domain including the arbitrary time $t_0$.
Indeed, this has rigorously been demonstrated to be the case
\cite{kuni95, ku97, ei20}. 
Now, the envelope curve can be constructed using the following envelope equation \cite{kuni95}:
\begin{equation}
\left. 
\frac{dx(t;t_{0})}{dt_{0}}
\right| _{t_{0}=t}
=\left.
\frac{\partial x}{\partial t_{0}}
\right| _{t_{0}=t}
+\left.
\frac{dA(t_{0})}{dt_{0}}\frac{\partial x}{\partial A}
\right| _{t_{0}=t}
+\left.
\frac{d\theta (t_{0})}{dt_{0}}\frac{\partial x}{\partial \theta}
\right| _{t_{0}=t} 
=0.
\label{eqR11}
\end{equation}
Note that we have 
taken into account the fact that
the integral constants $A$ and $\theta$ depend on the `initial time' $t=t_{0}$, and \eqref{eqR11} actually gives the dynamical equations for these variables. As will be done shortly, the insertion of the solutions to the dynamic equation into \eqref{eqR10} gives 
an approximate but globally valid solution to the original equation.
Because the envelope equation \eqref{eqR11} takes a similar form 
as the RG equation in quantum field theory,
it is also called the RG equation, and the asymptotic/global
analysis based on this equation was named the RG method \cite{golde89}.

Substituting Eq. \eqref{eqR10} into Eq. \eqref{eqR11}, we have

\begin{align}
0=&\varepsilon\left\{
\frac{dA}{dt}-\varepsilon^{2}f_{1a}A^{3}-\varepsilon (d_{1}+\varepsilon f_{1b})A
\right\} \cos(\omega_{0} t+\theta) \notag \\
&+\varepsilon A\left\{
-\frac{d\theta}{dt}-\varepsilon^{2} f_{2a}A^{2}-\varepsilon (d_{2}+\varepsilon f_{2b})
\right\} \sin(\omega_{0} t+\theta) \notag \\
&+\varepsilon^{2}A\left\{
2(d_{3}+\varepsilon f_{5})\frac{dA}{dt}+2(d_{4}+\varepsilon f_{6})A\frac{d\theta}{dt}-\varepsilon f_{7}A
\right\} \cos(2(\omega_{0}t+\theta)) \notag \\
&+\varepsilon^{2}A\left\{
2(d_{4}+\varepsilon f_{6})\frac{dA}{dt}-2(d_{3}+\varepsilon f_{5})A\frac{d\theta}{dt}-\varepsilon f_{8}A
\right\} \sin(2(\omega_{0}t+\theta)) \notag \\
&+3\varepsilon^{3}A^{2}\left\{
f_{9}\frac{dA}{dt}+f_{10}A\frac{d\theta}{dt}
\right\} \cos(3(\omega_{0}t+\theta))
+3\varepsilon^{3}A^{2}\left\{
f_{10}\frac{dA}{dt}-f_{9}A\frac{d\theta}{dt}
\right\} \sin(3(\omega_{0}t+\theta)) \notag \\
&+2\varepsilon^{2}(d_{5}+\varepsilon f_{11})A\frac{dA}{dt}
-\varepsilon^{3} f_{12}A+o(\varepsilon^{3}).
\label{eqR12}
\end{align}

Because $dA/dt$ and $d\theta/dt$ are of order $\varepsilon$, the coefficients $\cos(2(\omega_{0}t+\theta))$, 
$\sin(2(\omega_{0}+\theta))$, 
$\cos(3(\omega_{0}t+\theta))$, $\sin(3(\omega_{0}t+\theta))$, and $AdA/dt$ are of order $\varepsilon^{3}$ or 
higher. 
To make Eq. \eqref{eqR12} hold for any $t$, 
we only must ensure that 
the coefficients of the independent functions, namely $\cos(\omega_{0}t+\theta)$ and $\sin(\omega_{0}t+\theta)$,
vanish, and hence, we have 

\begin{align}
&\frac{dA}{dt}=\varepsilon^{2}f_{1a}A^{3}+\varepsilon(d_{1}+\varepsilon f_{1b})A+o(\varepsilon^{2}),  
\label{eqR13} \\
&\frac{d\theta}{dt}=-\varepsilon^{2}f_{2a}A^{2}-\varepsilon (d_{2}+\varepsilon f_{2b})+o(\varepsilon^{2}),
\label{eqR88}
\end{align}
which are the dynamic equations governing the 'integral constants' $A$ and $\theta$. We now see that
the integral constants have been lifted to dynamic variables 
through the RG/envelope equation.
The amplitude equation \eqref{eqR13} can be readily solved analytically. 
For instance, when $f_{1a}\,<\,0$ and $d_1+\varepsilon f_{1b}\,>\,0$, 
it yields 
\begin{equation}
A(t)=A_0\frac{{\cal A}}{\sqrt{{\cal A}^2+(A_0^2-{\cal A}^2){e}^{-2\alpha t}}},
\label{eqR_ampsol}
\end{equation}
where 
$\alpha=\varepsilon(d_1+\varepsilon f_{1b})$ and 
\begin{equation}
A_{0}=\sqrt{-\frac{d_{1}+\varepsilon f_{1b}}{\varepsilon f_{1a}}},
\label{eqR89}
\end{equation}
with ${\cal A}$ being the initial amplitude.
Equation \eqref{eqR_ampsol} indicates that the amplitude approaches $A_0$ monotonically as 
$t\, \to \, \infty$,
implying that $A_0$ is nothing but the amplitude of the limit cycle admitted 
in the original equation \eqref{eqR1}. 
Furthermore, Eq. \eqref{eqR88} indicates that the angular frequency on the limit cycle reads
\begin{equation}
\omega=\omega_{0}+(d\theta/dt)|_{A=A_{0}}
=
\omega_{0}-\varepsilon\frac{d_{2}f_{1a}-d_{1}f_{2a}}{f_{1a}},
\label{eqR90}
\end{equation}
which is constant.
The globally valid solution is given as the envelope of the 
family of curves, as previously stated. 
Thus, the solution on the limit cycle, which valid in a global
domain in the asymptotic regime, reads 
\begin{align}
x(t)=&x(t;t_{0})|_{t_{0}=t} \notag \\
=&x_0+\varepsilon A_{0}\cos(\omega t+\theta_{0})
+\varepsilon^{2}\{ d_{3}A_{0}^{2}\cos(2(\omega t+\theta_{0}))
+d_{4}A_{0}^{2}\sin(2(\omega t+\theta_{0}))+d_{5}A_{0}^{2}\} \notag \\
&+\varepsilon^{3}\{ f_{5}A_{0}^{2}\cos(2(\omega t+\theta_{0}))
+f_{6}A_{0}^{2}\sin(2(\omega t+\theta_{0}))
+f_{9}A_{0}^{3}\cos(3(\omega t+\theta_{0})) \notag \\
&+f_{10}A_{0}^{3}\sin(3(\omega t+\theta_{0}))
+f_{11}A_{0}^{2}
\} +o(\varepsilon^{3}).
\label{eqR91}
\end{align}

The RG method is a powerful method for obtaining a globally valid solution. 
This method can be applied to various models including discrete, stochastic, and partial differential equations, 
as given in \citep{kuni22, kuni98}.

\subsection*{A.2 Derivation of a time-evolution solution in a circadian rhythm model using the RG method}
\label{App5_2}
In this subsection, we applied the RG method to derive an approximate but globally valid solution of 
the 
circadian clock model given by Eq. \eqref{eq:G1}-\eqref{eq:G3}, which is
 a system of first-order equations with three variables.
We first convert the system into a single equation with higher-order derivatives as

\begin{equation}
\frac{d^{3}x_{3}}{dt^{3}}
+s_{1}\frac{d^{2}x_{3}}{dt^{2}}
+s_{2}\frac{dx_{3}}{dt}
+s_{3}x_{3}=p_{1}p_{2}f(x_{3})
\label{eqR17}
\end{equation}

where $s_{1}=k_{1}+k_{2}+k_{3}$, $s_{2}=k_{1}k_{2}+k_{2}k_{3}+k_{3}k_{1}$, $s_{3}=k_{1}k_{2}k_{3}$. 
To obtain the approximate solution, we set the transcriptional regulator $f(x_{3})$ as $r/x_{3}^{n}$. 
This model has a fixed point

\begin{equation}
x_{0}=\left( \frac{s_{3}}{p_{1}p_{2}r} \right) ^{-\frac{1}{n+1}},
\label{eqR18}
\end{equation}
which is destabilized through Hopf bifurcation for
\begin{equation}
n> n_{0}\equiv \frac{s_{4}}{s_{3}},
\label{eqR20}
\end{equation}
with $s_4=(k_1+k_2)(k_2+k_3)(k_3+k_1)$.

Then, we expand the solution around $t=t_{0}$ as a series
of $\varepsilon$

\begin{equation}
x_{3}(t;t_{0}) = x_{0}+\varepsilon u_{1}(t;t_{0})+\varepsilon^{2} u_{2}(t;t_{0})+\varepsilon ^{3}u_{3}(t;t_{0}) +O(\varepsilon^{3}). 
\label{eqR21}
\end{equation}

Substituting Eq. \eqref{eqR21} into Eq. \eqref{eqR17} and
equating the coefficients with the same powers of $\varepsilon$, we obtain

\begin{align}
&O(\varepsilon): \frac{d^{3}u_{1}}{dt^{3}}+s_{1}\frac{d^{2}u_{1}}{dt^{2}}
+s_{2}\frac{du_{1}}{dt}+s_{1}s_{2}u_{1}=0,
\label{eqR22} \\
&O(\varepsilon^{2}): \frac{d^{3}u_{2}}{dt^{3}}+s_{1}\frac{d^{2}u_{2}}{dt^{2}}
+s_{2}\frac{du_{2}}{dt}+s_{1}s_{2}u_{2}
=-s_{3}u_{1}+B_{1}u_{1}^{2},
\label{eqR23} \\
&O(\varepsilon^{3}): \frac{d^{3}u_{3}}{dt^{3}}+s_{1}\frac{d^{2}u_{3}}{dt^{2}}
+s_{2}\frac{du_{3}}{dt}+s_{1}s_{2}u_{3}
=-s_{3}u_{2}+B_{2}u_{1}u_{2}+B_{3}u_{1}^{2}+B_{4}u_{1}^{3}, 
\label{eqR24}
\end{align}
where

\begin{align}
&B_{1}=\frac{s_{4}s_{2}s_{1}}{2s_{3}}
\left( \frac{s_{3}}{p_{1}p_{2}r} \right) ^{\frac{s_{3}}{s_{1}s_{2}}},
\label{eqR27} \\
&B_{2}=\frac{s_{4}s_{2}s_{1}}{s_{3}}
\left( \frac{s_{3}}{p_{1}p_{2}r} \right) ^{\frac{s_{3}}{s_{1}s_{2}}},
\label{eqR28} \\
&B_{3}=\left\{
s_{4}+\frac{s_{3}}{2}-\left( \frac{s_{4}s_{3}}{2s_{2}s_{1}}
\ln\left( \frac{s_{3}}{p_{1}p_{2}r} \right)
\right)
\right\}
\left( \frac{s_{3}}{p_{1}p_{2}r} \right) ^{\frac{s_{3}}{s_{1}s_{2}}},
\label{eqR29} \\
&B_{4}=\frac{(s_{2}s_{1}+s_{3})s_{4}s_{1}s_{2}}{6s_{3}^2}
\left( \frac{2s_{3}}{p_{1}p_{2}r} \right) ^{\frac{s_{3}}{s_{1}s_{2}}}.
\label{eqR30}
\end{align}

Because Eq. \eqref{eqR22} has three eigenvalues, namely $\lambda_{1,2}=\pm i\sqrt{s_{2}}$ and $\lambda_{3}=-s_{1}$, the general solution of Eq. \eqref{eqR22} reads

\begin{equation}
u_{1}(t;t_{0})=A(t_{0})\cos(\omega_{0}t+\theta (t_{0}))+c(t_{0})e^{-s_{1}t}, \quad \quad (\omega_0:= \sqrt{s_2}),
\label{eqR31}
\end{equation}
where $A$, $\theta$, and $c$ are integral constants that depend on initial time $t_{0}$.

Considering the asymptotic regime in which the second term in Eq. \eqref{eqR31} is so small and negligible,
then the first-order solution can be written as

\begin{equation}
u_{1}(t;t_{0})=A(t_{0})\cos(\omega_{0}t+\theta (t_0)).
\label{eqR32}
\end{equation}

Substituting Eq. \eqref{eqR32} into Eq. \eqref{eqR23}, we obtain

\begin{equation}
\frac{d^{3}u_{2}}{dt^{3}}+s_{1}\frac{d^{2}u_{2}}{dt^{2}}
+s_{2}\frac{du_{2}}{dt}+s_{1}s_{2}u_{2}
=-s_{3}A\cos(\omega_{0}t+\theta)+CA^{2}\cos(2\omega_{0}t+2\theta)
+CA^{2},
\label{eqR33}
\end{equation}

where

\begin{equation}
C=\frac{s_{4}s_{1}s_{2}}{4s_{3}}
\left( \frac{s_{3}}{p_{1}p_{2}r} \right) ^{\frac{s_{3}}{s_{1}s_{2}}}.
\label{eqR34}
\end{equation}

Equation \eqref{eqR33} is an inhomogeneous equation that contains a zero mode of the linear operator, 
and the solution 
having a suitable form for applying the RG method is written as
\begin{align}
u_{2}(t;t_{0})=
&D_{1}A(t-t_{0})\cos(\omega_{0}t+\theta)
-D_{2}A(t-t_{0})\sin(\omega_{0}t+\theta) \notag \\
&-D_{3}A^{2}\cos(2\omega_{0}t+2\theta)-D_{4}A^{2}\sin(2\omega_{0}t+2\theta)
+D_{5}A^{2},
\label{eqR35}
\end{align}

where the coefficients are

\begin{align}
&D_{1}=\frac{s_{3}}{2(s_{1}^2+s_{2})},
\label{eqR36} \\
&D_{2}=\frac{s_{1}s_{3}}{2\sqrt{s_{2}}(s_{1}^2+s_{2})},
\label{eqR37} \\
&D_{3}=\frac{s_{4}s_{1}^2}{12(s_{1}^2+4s_{2})s_{3}}
\left( \frac{s_{3}}{p_{1}p_{2}r} \right) ^{\frac{s_{3}}{s_{1}s_{2}}},
\label{eqR38} \\
&D_{4}=\frac{s_{4}s_{1}\sqrt{s_{2}}}{6(s_{1}^2+4s_{2})s_{3}}
\left( \frac{s_{3}}{p_{1}p_{2}r} \right) ^{\frac{s_{3}}{s_{1}s_{2}}},
\label{eqR39} \\
&D_{5}=\frac{s_{4}}{4s_{3}}
\left( \frac{s_{3}}{p_{1}p_{2}r} \right) ^{\frac{s_{3}}{s_{1}s_{2}}}.
\label{eqR40}
\end{align}

Tentatively, after collecting all the obtained terms, we have the approximate solution in the second order as

\begin{align}
x_{3}(t;t_{0})=&x_{0}+\varepsilon A\cos(\omega_{0}t+\theta)
+\varepsilon^{2}\{ D_{1}A(t-t_{0})\cos(\omega_{0}t+\theta)
-D_{2}A(t-t_{0})\sin(\omega_{0}t+\theta) \notag \\
&-D_{3}A^{2}\cos(2\omega_{0}t+2\theta)-D_{4}A^{2}\sin(2\omega_{0}t+2\theta)
+D_{5}A^{2}\} +o(\varepsilon^{2}).
\label{eqR41}
\end{align}

Because it contains the secular terms, the solution diverges as $|t-t_{0}|$ 
goes infinity.
To resum the would-be divergent terms, 
we apply the RG equation $dx_3(t, t_0)/dt_0\vert_{t_0=t}$, 
which leads to the equations governing the amplitude and phase as
\[
dA/dt=\varepsilon s_{3}A/2(s_{1}^{2}+s_{2}),
\]
and 
\[
d\theta /dt=\varepsilon s_{1}s_{3}/2\sqrt{s_{2}}(s_{1}^{2}+s_{2}),
\]
nicely describing the slow motions of the amplitude and phase, respectively.
However, it fails to describe a transitional behavior approaching a limit cycle, as indicated by the present model.

Therefore, we analyzed the third-order equation, which might lead to 
a limit cycle solution. 
Substituting Eqs. \eqref{eqR32} and \eqref{eqR35} into Eq. \eqref{eqR24}, we have

\begin{align}
&\frac{d^{3}u_{3}}{dt^{3}}+s_{1}\frac{d^{2}u_{3}}{dt^{2}}
+s_{2}\frac{du_{3}}{dt}+s_{1}s_{2}u_{3}
=
E_{1}A^{3}\cos(\omega_{0}t+\theta)-E_{2}A^{3}\sin(\omega_{0}t+\theta)
\notag \\
&-E_{3}A(t-t_{0})\cos(\omega_{0}t+\theta)
+E_{4}A(t-t_{0})\sin(\omega_{0}t+\theta)
+E_{5}A^{2}\cos(2\omega_{0}t+2\theta)
\notag \\
&+E_{6}A^{2}\sin(2\omega_{0}t+2\theta)
+E_{7}A^{2}(t-t_{0})\cos(2\omega_{0}t+2\theta)
-E_{8}A^{2}(t-t_{0})\sin(2\omega_{0}t+2\theta)
\notag \\
&-E_{9}A^{3}\cos(3\omega_{0}t+3\theta)
-E_{10}A^{3}\sin(3\omega_{0}t+3\theta)
+E_{11}A^{2}+E_{12}A^{2}(t-t_{0}).
\label{eqR44}
\end{align}

where

\begin{align}
&E_{1}=\frac{(s_{1}^{2}s_{2}-4s_{1}^{2}s_{3}+6s_{1}s_{2}^{2}-18s_{2}s_{3})s_{4}s_{1}s_{2}}
{12(s_{1}^{2}+4s_{2})s_{3}^{2}}
\left( \frac{s_{3}}{p_{1}p_{2}r} \right) ^{\frac{2s_{3}}{s_{1}s_{2}}},
\label{eqR52} \\
&E_{2}=\frac{s_{4}^{2}s_{1}^{2}\sqrt{s_{2}^{3}}}
{12(s_{1}^{2}+4s_{2})s_{3}^{2}}
\left( \frac{s_{3}}{p_{1}p_{2}r} \right) ^{\frac{2s_{3}}{s_{1}s_{2}}},
\label{eqR53} \\
&E_{3}=\frac{s_{3}^{2}}{2(s_{1}^{2}+s_{2})},
\label{eqR54} \\
&E_{4}=\frac{s_{1}s_{3}^{2}}{2(s_{1}^{2}+s_{2})\sqrt{s_{2}}},
\label{eqR55} \\ 
&E_{5}=\left\{
\frac{7s_{1}^{3}s_{2}-4s_{1}^{2}s_{3}+24s_{1}s_{2}^{2}-12s_{2}s_{3}}
{12(s_{1}^{2}+4s_{2})}
-\frac{s_{4}s_{3}}{4s_{1}s_{2}}\ln\left( \frac{s_{3}}{p_{1}p_{2}r} \right)
\right\}
\left( \frac{s_{3}}{p_{1}p_{2}r} \right) ^{\frac{s_{3}}{s_{1}s_{2}}},
\label{eqR56} \\
&E_{6}=\frac{s_{4}s_{1}\sqrt{s_{2}}}
{6(s_{1}^{2}+4s_{2})}
\left( \frac{s_{3}}{p_{1}p_{2}r} \right) ^{\frac{s_{3}}{s_{1}s_{2}}},
\label{eqR57} \\
&E_{7}=\frac{s_{4}s_{1}s_{2}}
{4(s_{1}^{2}+s_{2})}
\left( \frac{s_{3}}{p_{1}p_{2}r} \right) ^{\frac{s_{3}}{s_{1}s_{2}}},
\label{eqR58} \\
&E_{8}=\frac{s_{4}s_{1}^{2}\sqrt{s_{2}}}
{4(s_{1}^{2}+s_{2})}
\left( \frac{s_{3}}{p_{1}p_{2}r} \right) ^{\frac{s_{3}}{s_{1}s_{2}}},
\label{eqR59} \\
&E_{9}=\frac{((s_{1}^2+2s_{2})s_{1}+2s_{3})s_{4}s_{1}s_{2}^{2}}
{12(s_{1}^{2}+4s_{2})s_{3}^{2}}
\left( \frac{s_{3}}{p_{1}p_{2}r} \right) ^{\frac{2s_{3}}{s_{1}s_{2}}},
\label{eqR60} \\
&E_{10}=\frac{s_{4}^{2}s_{1}^{2}\sqrt{s_{2}^{3}}}
{12(s_{1}^{2}+4s_{2})s_{3}^{2}}
\left( \frac{s_{3}}{p_{1}p_{2}r} \right) ^{\frac{2s_{3}}{s_{1}s_{2}}},
\label{eqR61} \\
&E_{11}=\frac{1}{4}\left\{
s_{1}s_{2}
-\frac{s_{4}s_{3}}{s_{1}s_{2}}
\ln\left( \frac{s_{3}}{p_{1}p_{2}r} \right)
\right\}
\left( \frac{s_{3}}{p_{1}p_{2}r} \right) ^{\frac{2s_{3}}{s_{1}s_{2}}},
\label{eqR62} \\
&E_{12}=\frac{s_{4}s_{1}s_{2}}
{4(s_{1}^{2}+s_{2})}
\left( \frac{s_{3}}{p_{1}p_{2}r} \right) ^{\frac{s_{3}}{s_{1}s_{2}}}.
\label{eqR63}
\end{align}

The solution to Eq. \eqref{eqR44} 
is given by 
\begin{align}
u_{3}(t;t_{0})=&-(F_{1a}A^{3}+F_{1b}A)(t-t_{0})\cos(\omega_{0}t+\theta)
+(F_{2a}A^{3}+F_{2b}A)(t-t_{0})\sin(\omega_{0}+\theta)
\notag \\
&-F_{3}A(t-t_{0})^{2} \cos(\omega_{0}t+\theta)
-F_{4}A(t-t_{0})^{2} \sin(\omega_{0}t+\theta)
+F_{5}A^{2}\cos(2\omega_{0}t+2\theta)
\notag \\
&+F_{6}A^{2}\sin(2\omega_{0}t+2\theta)
-F_{7}A^{2}(t-t_{0})\cos(2\omega_{0}t+2\theta)
\notag \\
&+F_{8}A^{2}(t-t_{0})\sin(2\omega_{0}t+2\theta)
+F_{9}A^{3}\cos(3\omega_{0}t+3\theta)
+F_{10}A^{3}\sin(3\omega_{0}t+3\theta)
\notag \\
&+F_{11}A^{2}+F_{12}A^{2}(t-t_{0})
\label{eqR64}
\end{align}
where
\begin{align}
&F_{1a}=\frac{(2s_{1}s_{2}^{2}-(s_{1}^{2}+6s_{2})s_{3})s_{4}s_{1}s_{2}}
{8(s_{1}^{2}+4s_{2})(s_{1}^{2}+s_{2})s_{3}^{2}}
\left( \frac{s_{3}}{p_{1}p_{2}r} \right) ^{\frac{2s_{3}}{s_{1}s_{2}}},
\label{eqR65} \\
&F_{1b}=\frac{s_{1}s_{3}^{2}}
{(s_{1}^{2}+s_{2})^{3}},
\label{eqR66} \\
&F_{2a}=\frac{((s_{1}^{2}+7s_{2})s_{1}s_{2}-(4s_{1}^{2}+19s_{2})s_{3})s_{4}s_{1}^{2}\sqrt{s_{2}}}
{24(s_{1}^{2}+s_{2})(s_{1}^{2}+4s_{2})s_{3}^{2}}
\left( \frac{s_{3}}{p_{1}p_{2}r} \right) ^{\frac{2s_{3}}{s_{1}s_{2}}},
\label{eqR67} \\
&F_{2b}=\frac{((s_{1}^{2}+6s_{2})s_{1}^{2}-3s_{2}^{2})s_{3}^{2}}
{8(s_{1}^{2}+s_{2})^{3}\sqrt{s_{2}^{3}}},
\label{eqR68} \\
&F_{3}=\frac{(s_{1}^{2}-s_{2})s_{3}^{2}}{8(s_{1}^{2}+s_{2})s_{2}},
\label{eqR69} \\
&F_{4}=\frac{s_{1}s_{3}^{2}}{4(s_{1}^{2}+s_{2})\sqrt{s_{2}}},
\label{eqR70} \\
&F_{5}=\left\{
\frac{s_{4}s_{3}}{12(s_{1}^{2}+4s_{2})s_{2}^{2}}
\ln\left( \frac{s_{3}}{p_{1}p_{2}r} \right) \right.
\notag \\
&\hspace{10mm} \left. -
\frac{(3s_{1}^{5}+4s_{1}^{3}s_{2}+11s_{1}^{2}s_{3}+64s_{1}s_{2}^{2}-52s_{2}s_{3})s_{1}}
{36(s_{1}^{2}+s_{2})(s_{1}^{2}+4s_{2})^{2}}
\right\}
\left( \frac{s_{3}}{p_{1}p_{2}r} \right) ^{\frac{s_{3}}{s_{1}s_{2}}},
\label{eqR71} \\
&F_{6}=\left\{
\frac{s_{4}s_{3}}{6(s_{1}^{2}+4s_{2})s_{1}\sqrt{s_{2}^{3}}}
\ln\left( \frac{s_{3}}{p_{1}p_{2}r} \right) \right.
\notag \\
&\hspace{10mm} \left. -
\frac{7s_{1}^{5}s_{2}-s_{1}^{4}s_{3}-8s_{1}^{3}s_{2}^{2}+38s_{1}^{2}s_{2}s_{3}+48s_{1}s_{2}^{3}-24s_{2}^{2}s_{3}}
{36(s_{1}^{2}+s_{2})(s_{1}^{2}+4s_{2})^{2}\sqrt{s_{2}}}
\right\}
\left( \frac{s_{3}}{p_{1}p_{2}r} \right) ^{\frac{s_{3}}{s_{1}s_{2}}},
\label{eqR72} \\
&F_{7}=\frac{s_{4}s_{1}^{2}}{4(s_{1}^{2}+s_{2})(s_{1}^{2}+4s_{2})}
\left( \frac{s_{3}}{p_{1}p_{2}r} \right) ^{\frac{s_{3}}{s_{1}s_{2}}},
\label{eqR73} \\
&F_{8}=\frac{(s_{1}^{2}-2s_{2})s_{4}s_{1}}
{12(s_{1}^{2}+s_{2})(s_{1}^{2}+4s_{2})\sqrt{s_{2}}}
\left( \frac{s_{3}}{p_{1}p_{2}r} \right) ^{\frac{s_{3}}{s_{1}s_{2}}},
\label{eqR74} \\
&F_{9}=\frac{(s_{1}(s_{1}^{2}-s_{2})+5s_{3})s_{4}s_{1}^{2}s_{2}}
{96(s_{1}^{2}+4s_{2})(s_{1}^{2}+9s_{2})s_{3}^{2}}
\left( \frac{s_{3}}{p_{1}p_{2}r} \right) ^{\frac{2s_{3}}{s_{1}s_{2}}},
\label{eqR75} \\
&F_{10}=\frac{(2(2s_{1}^{2}+3s_{2})s_{1}s_{2}+(6s_{2}-s_{1}^{2})s_{3})s_{4}s_{1}\sqrt{s_{2}}}
{96(s_{1}^{2}+4s_{2})(s_{1}^{2}+9s_{2})s_{3}^{2}}
\left( \frac{s_{3}}{p_{1}p_{2}r} \right) ^{\frac{2s_{3}}{s_{1}s_{2}}},
\label{eqR76} \\
&F_{11}=\left\{
\frac{(s_{1}^{3}+s_{3})}{4(s_{1}^{2}+s_{2})s_{1}}
-\frac{s_{4}s_{3}}{4s_{1}^{2}s_{2}^{2}}
\ln\left( \frac{s_{3}}{p_{1}p_{2}r} \right) 
\right\}
\left( \frac{s_{3}}{p_{1}p_{2}r} \right) ^{\frac{s_{3}}{s_{1}s_{2}}},
\label{eqR77} \\
&F_{12}=\frac{s_{4}}{4(s_{1}^{2}+s_{2})}
\left( \frac{s_{3}}{p_{1}p_{2}r} \right) ^{\frac{s_{3}}{s_{1}s_{2}}}.
\label{eqR78}
\end{align}

To obtain a globally valid solution using Eq. \eqref{eqR64}, we apply
the RG method, which utilizes the RG equation
\begin{align}
\left. \frac{dx_{3}(t;t_{0})}{dt_{0}} \right| _{t_{0}=t}
=& \left. \frac{\partial x_{3}(t;t_{0})}{\partial t_{0}} \right| _{t_{0}=t}
+\left. \frac{dA}{dt_{0}}\frac{\partial x_{3}(t;t_{0})}{\partial A} \right| _{t_{0}=t}
+\left. \frac{d\theta}{dt_{0}}\frac{\partial x_{3}(t;t_{0})}{\partial \theta} \right| _{t_{0}=t}
\notag \\
=&\left\{ \varepsilon \frac{dA}{dt}-\varepsilon D_{1}A+\varepsilon ^{3}(F_{1a}A^{3}+F_{1b}A) \right\} \cos(\omega_{0}t+\theta)
\notag \\
&+\left\{ -\varepsilon A\frac{d\theta}{dt}+\varepsilon ^{2}D_{2}A-\varepsilon ^{3}(F_{2a}A^{3}+F_{2b}A) \right\} \sin(\omega_{0}t+\theta)
=0,
\label{eqR79}
\end{align}
where we have neglected the higher-order terms $o(\varepsilon^{3})$.
For Eq. \eqref{eqR79} to hold for any $t$, the coefficients of the two independent functions should vanish. Thus, we obtain the dynamic equations for $A$ and $\theta$:
\begin{align}
&\frac{dA}{dt}=\varepsilon D_{1}A-\varepsilon^{2}(F_{1a}A^{3}+F_{1b}A)
+o(\varepsilon^{2}),
\label{eqR80} \\
&\frac{d\theta}{dt}=\varepsilon D_{2}-\varepsilon^{2}(F_{2a}A^{2}+F_{2b})
+o(\varepsilon^{2}).
\label{eqR81}
\end{align}
The amplitude equation \eqref{eqR80} has a new fixed point 
\begin{equation}
A_{0}=\sqrt{\frac{D_{1}-\varepsilon F_{1b}}{\varepsilon F_{1a}}}
=\sqrt{\frac{4((s_{1}^{2}+s_{2})^{2}-2\varepsilon s_{1}s_{3})(s_{1}^{2}+4s_{2})s_{3}^{3}}
{\varepsilon (2s_{1}s_{2}^{2}-(s_{1}^{2}+6s_{2})s_{3})(s_{1}^{2}+s_{2})^{2}s_{4}s_{1}s_{2}}}
\left( \frac{p_{1}p_{2}r}{s_{3}} \right) ^{\frac{s_{3}}{s_{1}s_{2}}},
\label{eqR82}
\end{equation}
which is 
nothing but the amplitude of the desired limit cycle.
The phase function $\theta(t)$ on the limit cycle is expressed as 
\begin{equation}
\theta (t)=(\varepsilon D_{2}-\varepsilon^{2}(F_{2a}A_{0}^{2}+F_{2b})+o(\varepsilon^{2}))t+\theta _{0},
\label{eqR83}
\end{equation}
where $\theta _{0}$ is the integral constant and it gives 
the initial phase at $t=0$. 
Substituting Eqs. \eqref{eqR37}, \eqref{eqR67}, \eqref{eqR68} and \eqref{eqR82} into \eqref{eqR83}, $\theta (t)$ is reduced to
\begin{equation}
\theta (t)=\left\{ 
-\varepsilon\frac{s_{1}s_{3}s_{4}}{6(2s_{1}s_{2}^{2}-(s_{1}^{2}+6s_{2})s_{3})\sqrt{s_{2}}}+o(\varepsilon)
\right\}
t +\theta _{0}.
\label{eqR85}
\end{equation}
Thus, the solution describing the limit cycle, which is valid in a global domain in the asymptotic regime, is given by
\begin{align}
x_{3}(t)=&x_{3}(t;t_{0})|_{t_{0}=t}
\notag \\
=&x_{0}+\varepsilon A_{0}\cos(\omega t+\theta _{0})
\notag \\
&-\varepsilon^{2}\{ D_{3}A_{0}^{2}\cos(2(\omega t+\theta_{0}))
+D_{4}A_{0}^{2}\sin(2(\omega t+\theta_{0}))
\} +o(\varepsilon ^{2}).
\label{eqR86}
\end{align}
In particular, if the initial phase is set to be $\theta_{0} =-\pi /2$, we have Eq. \eqref{eqR45}.

\subsection*{A.3 Numerical and 
 RG analyses of the Lotka-Volterra model}

Equation \eqref{eq:tau} and the numerical simulation indicate that non-sinusoidal power ($NS$) tends to be larger when the period
hardly changes and is stable in response to increases of the parameter values specifying the degradation rates, as presented in Fig. \ref{figure3}B. This suggests that the waveform becomes more distorted at higher temperatures when the circadian period is temperature-compensated. 
It was previously reported that the same conclusion holds for other oscillatory models, including a realistic mammalian circadian clock model, a post-translational model in cyanobacteria, and the van der Pol oscillator \cite{kuro19, gibo20}. 
However, it is important to note that the findings using specific mathematical models might not be universally applicable to other models and actual organisms. 
Therefore, one should examine whether the waveform also plays a crucial role in the stability of the period in other oscillatory models. 

\begin{figure}[H]
\begin{center}	
 \hspace*{0cm}
 \includegraphics[width=16cm]{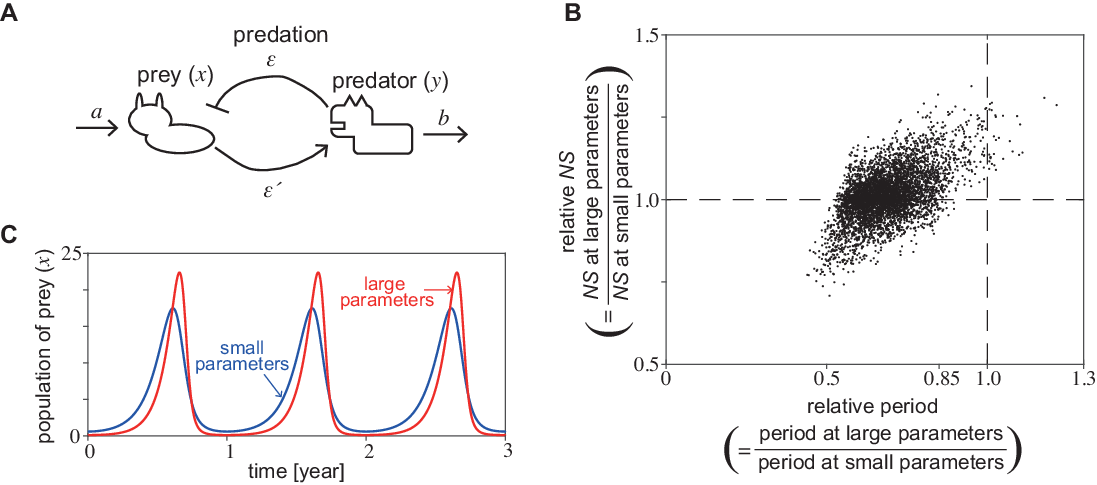}
 \caption{(A) Lotka-Volterra model. 
 (B) Distribution of relative $NS$ for $x(t)$ of the Lotka-Volterra model as a function of the relative period when rate constants are increased. We first generated reference parameter sets. Then, each parameter $a$, $b$, $\varepsilon$, and $\epsilon '$ in the model's reference parameter set was randomly multiplied by a factor of $1.1$-$1.9$. (C) Examples of the waveform in the Lotka-Volterra model when the period is relatively unchanged. The blue and red lines represent small and large parameters, respectively.}	
 \label{figure9}	
 \end{center}
\end{figure}

Thus, we conduct a numerical simulation to test the possible period-waveform correlation 
in the Lotka-Volterra model as done for the circadian clock model.
Needless to say, the Lotka-Volterra model is one of the most extensively studied mathematical models in biology \cite{murray88}, and it effectively explains population dynamics in prey-predator systems.
The Lotka-Volterra model is given as a system with two variables as
\begin{align}
&\frac{dx}{dt}=ax-\varepsilon xy,
\label{eq2} \\
&\frac{dy}{dt}=-by+\varepsilon 'xy,
\label{eq3}
\end{align}
where $x(t)$ and $y(t)$ are numbers of prey and predators. Parameter $a$ is the growth rate of prey, $b$ is the death rate of predators, $\varepsilon$ is the death rate of prey attributable to predation, and $\varepsilon '$ is the growth rate of predators (Supplementary Fig. \ref{figure9}A). 

In the present numerical simulation, we first generated
$100$ reference parameter sets corresponding to the reference temperature.
The values of the model parameters $a$ and $b$ were generated randomly with a uniform distribution between $0$ and $1$, and similarly, the values of the other model parameters $\varepsilon$ and $\varepsilon '$ 
were also randomly assigned values between $0$ and $0.5$. The period obtained for each parameter set with the initial condition $(x_{0},y_{0})=(2,2)$ was denoted as $\tau_1$. 
Next, $a$, $b$, $\varepsilon$, and $\varepsilon '$ 
were multiplied by a random factor within the range of $1.1$-$1.9$ to simulate the increase in temperature, yielding $49$ oscillatory parameter sets. Each resulting new period was denoted as $\tau_2$, and thus, the ratio of the period $=\tau_2/\tau_1$, which is called the relative period,
was obtained.

The numerical simulations reveal a consistent positive correlation between $NS$ and the relative period (Supplementary Fig. \ref{figure9}BC). 
A notable point is that the value of $NS$ tends to increase along with the period when larger multiplicative factors are used in the simulation. 
This result again suggests that our findings that the waveform becomes more distorted when the period remains relatively stable in response to increased parameter values
is a rather universal phenomenon not restricted to the behavior observed in the circadian clock model (Fig. \ref{figure7}A).

The period of the Lotka-Volterra model, together with its approximate but globally valid solution, was previously derived analytically by one of the present authors \cite{ku97, ei20} on the basis of the RG method; 
see also the pioneering work \cite{fra74} based on a different method. 
Next, we will demonstrate that the expression explicitly reveals that the period of the Lotka-Volterra model almost linearly increases with the waveform distortion for the average of the prey-predator time series (i.e. $\overline{NS}$). 
In the mathematical analysis, it proved convenient
to use new variables $(\xi(t), \eta(t))$ defined as
\[
x(t)=(b+\varepsilon\xi (t))/\varepsilon ' ,
\quad y(t)=a/\varepsilon +\eta (t).
\]
The RG method performed in \cite{kuni97, ei20}
in the second order of $\varepsilon$ leads to
\begin{align}
\xi(t)=&\left( 1-\varepsilon^{2}\frac{a-b}{4ab^{2}}\frac{A^{2}}{12} \right) A\sin\Theta
-\varepsilon^{2}\frac{1}{b\sqrt{ab}}\frac{A^{2}}{24}\cos\Theta 
-\varepsilon\frac{1}{\sqrt{ab}}\frac{A^{2}}{6}\sin(2\Theta) \notag \\
&-\varepsilon\frac{1}{b}\frac{A^{2}}{3}\cos(2\Theta)
-\varepsilon^{2}\frac{3a-b}{4ab^{2}}\frac{A^{3}}{8}\sin(3\Theta)
+\varepsilon^{2}\frac{1}{b\sqrt{ab}}\frac{A^{2}}{8}\cos(3\Theta) +o(\varepsilon^{2}),
\label{eq4} \\
\eta(t)=&\varepsilon^{2}\frac{1}{b^{2}}\frac{A^{3}}{24}\sin\Theta
-\frac{\sqrt{ab}}{b}\left( 1+\varepsilon^{2}\frac{a-b}{4ab^{2}}\frac{A^{2}}{12}\right) A\cos\Theta 
-\varepsilon\frac{\sqrt{ab}}{b^{2}}\frac{A^{2}}{6}\sin(2\Theta) \notag \\
&+\varepsilon\frac{1}{b}\frac{A^{2}}{3}\cos(2\Theta)
+\varepsilon^{2}\frac{1}{b^{2}}\frac{A^{3}}{8}\sin(3\Theta)
+\varepsilon^{2}\frac{a-3b}{4b^{2}\sqrt{ab}}\frac{A^{3}}{8}\cos(3\Theta) 
+o(\varepsilon^{2}),
\label{eq5}
\end{align}
where $A$ and $\theta$ are the integral constants, which
are to be determined by the initial condition,
and $\Theta=\tilde{\omega} t+\theta$, with
$\tilde{\omega}$ being the angular frequency given by

\begin{equation}
\tilde{\omega}=
\sqrt{ab} \left\{ 1-\frac{\varepsilon^{2}A^{2}(a+b)}{24ab^{2}}\right\},
\end{equation}
from which we have the formula of the period of the system after some manipulation as
\begin{equation}    
\tau=\frac{2\pi}{\tilde{\omega}}=\frac{2\pi}{\sqrt{ab}}\left\{ \frac{2}{5}\left( 1+\frac{5}{2}\frac{\varepsilon^{2}A^{2}(a+b)}{24ab^{2}} \right) +\frac{3}{5} \right\} +o(\varepsilon^{2}).
\label{eq10} 
\end{equation}

From the waveforms for $\xi(t)$ and $\eta(t)$ given by \eqref{eq4} and \eqref{eq5}, respectively, 
we can obtain the waveform distortion of each variable as follows:
\begin{align}
&NS^{(\xi)} =1+\frac{\varepsilon^{2}A^{2}(4a+b)}{24ab^{2}}+o(\varepsilon^{2}), 
\label{eqR117} \\
&NS^{(\eta)} =1+\frac{\varepsilon^{2}A^{2}(a+4b)}{24ab^{2}}+o(\varepsilon^{2}). 
\label{eqR118}
\end{align}

It is notable that the mean of $NS^{(\xi)}$ and $NS^{(\eta)}$ 
takes the form

\begin{equation}
\overline{NS}=\frac{1}{2}(NS^{(\xi)}+NS^{(\eta)})
=1+\frac{5}{2}\frac{\varepsilon^{2}A^{2}(a+b)}{24ab^{2}}+o(\varepsilon^{2}).
\label{eq9}
\end{equation}

Indeed, comparing \eqref{eq9} and \eqref{eq10}, we arrive at

\begin{equation}
\tau=\frac{2\pi}{\sqrt{ab}}\left(\frac{2}{5}\overline{NS}+\frac{3}{5} \right) 
+o(\varepsilon^{2}),
\end{equation}

which states that the period and the mean waveform distortion, namely $\overline{NS}$, are linearly dependent on each other, and they tend to increase (or decrease) in a parallel manner. This is what we aimed to demonstrate for the
Lotka-Volterra model.

\subsection*{A.4 
Numerical and RG analyses of the van der Pol model }
\label{App5_3}
We consider the van del Pol model as follows:
\begin{equation}
\frac{d^{2}x}{dt^{2}}+x=\varepsilon (1-x^{2})\frac{dx}{dt},
\label{eqR92}
\end{equation}

which is known as one of the fundamental non-linear oscillator models.
Previously, two of the authors derived the period formula $\tau =2\pi[\sum_{j=1}^{\infty}|a_{j}|^{2}j^{2}/\sum_{j=1}^{\infty}|a_{j}|^{2}]^{1/2}$, meaning that the period of the model is also proportional to $NS$ \cite{kuro19}. Then, we derive the approximate solution of the model using the RG method and confirm the proportionality between the period and the waveform distortion $NS$ in detail. First, we represent the local solution around $t=t_{0}$ as a perturbation series
\begin{equation}
x(t;t_{0})=x_{0}(t;t_{0})+\varepsilon x_{1}(t;t_{0})+\varepsilon^{2} x_{2}(t;t_{0})+o(\varepsilon^{2}).
\label{eqR93}
\end{equation}

Then, substituting Eq. \eqref{eqR93} into Eq. \eqref{eqR92} and equating the terms with the same powers of $\varepsilon$, we obtain

\begin{align}
&O(\varepsilon^{0}): \frac{d^{2}x_{0}}{dt^{2}}+x_{0}=0,
\label{eqR94} \\
&O(\varepsilon^{1}): \frac{d^{2}x_{1}}{dt^{2}}+x_{1}
=(1-x_{0}^{2})\frac{dx_{0}}{dt},
\label{eqR95} \\
&O(\varepsilon^{2}): \frac{d^{2}x_{2}}{dt^{2}}+x_{2}
=(1-x_{0}^{2})\frac{dx_{1}}{dt}-2x_{0}x_{1}\frac{dx_{0}}{dt}. 
\label{eqR96}
\end{align}

The solution for the zeroth-order equation \eqref{eqR94} is

\begin{equation}
x_{0}(t;t_{0})=A(t_{0})\cos(t+\theta(t_{0})),
\label{eqR97}
\end{equation}
where $A$ and $\theta$ are integral constants and they potentially depend on initial time $t_{0}$. Then, substituting Eq. \eqref{eqR97} into Eq. \eqref{eqR95}, we have

\begin{equation}
\frac{d^{2}x_{1}}{dt^{2}}+x_{1}
=-A\left( 1-\frac{A^{2}}{4} \right) \sin(t+\theta)+\frac{A^{3}}{4}\sin(3t+3\theta).
\label{eqR98}
\end{equation}
By solving the solution of Eq. \eqref{eqR98} around $t=t_{0}$, the first-order solution is given by

\begin{equation}
x_{1}(t;t_{0})
=\frac{A}{2}\left( 1-\frac{A^{2}}{4} \right) (t-t_{0})\cos(t+\theta)
-\frac{A^{3}}{32}\sin(3t+3\theta).
\label{eqR99}
\end{equation}
Similarly, by substituting zero-th and first-order solutions \eqref{eqR97} and \eqref{eqR99} into Eq. \eqref{eqR96}, the second-order equation is
\begin{align}
\frac{d^{2}x_{2}}{dt^{2}}+x_{2}
=&F_{1}(A)\cos(t+\theta)-F_{2}(A)(t-t_{0})\sin(t+\theta)+F_{3}(A)\cos(3t+3\theta)
\notag \\
&-F_{4}(A)(t-t_{0})\sin(3t+3\theta)+F_{5}(A)\cos(5t+5\theta),
\label{eqR100}
\end{align}
where 
\begin{align}
&F_{1}(A)=\frac{A}{2}\left( \frac{13}{64}A^{4}-A^{2}+1 \right) ,
\label{eqR101} \\
&F_{2}(A)=\frac{A}{2}\left( \frac{3}{16}A^{4}-A^{2}+1 \right),
\label{eqR102} \\
&F_{3}(A)=\frac{A^{3}}{32}\left( \frac{5}{2}A^{2}-7 \right),
\label{eqR103} \\
&F_{4}(A)=\frac{3A^{3}}{8}\left( \frac{1}{4}A^{2}-1 \right),
\label{eqR104} \\
&F_{5}(A)=\frac{5A^{5}}{128}.
\label{eqR105}
\end{align}
The solution of the second-order equation \eqref{eqR100} is
\begin{align}
x_{2}(t;t_{0})
=&\frac{1}{4}(2F_{1}(A)-F_{2}(A))(t-t_{0})\sin(t+\theta)
+\frac{1}{4}F_{2}(A)(t-t_{0})^{2}\cos(t+\theta) 
\notag \\
&+\frac{1}{32}(-4F_{3}(A)+3F_{4}(A))\cos(3t+3\theta)
+\frac{1}{8}F_{4}(A)(t-t_{0})\sin(3t+3\theta)
\notag \\
&-\frac{1}{24}F_{5}(A)\cos(5t+5\theta).
\label{eqR106}
\end{align}
Therefore, the perturbative solution up to the second-order of $\varepsilon$ is
\begin{align}
x(t;t_{0})
=&A\cos(t+\theta)
+\varepsilon\left\{
\frac{A}{2}\left( 1-\frac{A^{2}}{4} \right) (t-t_{0})\cos(t+\theta)
-\frac{A^{3}}{32}\sin(3t+3\theta)
\right\}
\notag \\
&+\varepsilon^{2}\left\{\frac{1}{4}(2F_{1}(A)-F_{2}(A))(t-t_{0})\sin(t+\theta)
+\frac{1}{4}F_{2}(A)(t-t_{0})^{2}\cos(t+\theta) \right.
\notag \\
&+\frac{1}{32}(-4F_{3}(A)+3F_{4}(A))\cos(3t+3\theta)
+\frac{1}{8}F_{4}(A)(t-t_{0})\sin(3t+3\theta)
\notag \\
&\left.
-\frac{1}{24}F_{5}(A)\cos(5t+5\theta).
\right\} +o(\varepsilon^{2}) .
\label{eqR107}
\end{align}
To obtain the globally valid solution using Eq. \eqref{eqR107}, we apply
the RG method, which utilizes the RG equation
\begin{align}
\left. \frac{dx}{dt_{0}} \right|_{t_{0}=t}=
&\left\{ \frac{dA}{dt}+\varepsilon \frac{1}{2}A\left( \frac{1}{4}A^{2}-1 \right) \right\} \cos(t+\theta)
\notag \\
&+\left\{ -A\frac{d\theta}{dt} -\varepsilon^{2}\frac{1}{4}(2F_{1}(A)-F_{2}(A)) \right\} \sin(t+\theta)=0,
\label{eqR108}
\end{align}
where we have neglected the higher-order terms $o(\varepsilon^{2})$.
For Eq. \eqref{eqR108} to hold for any $t$, the coefficients of the two independent functions should vanish. Thus, we end up with the 
dynamical equations for $A$ and $\theta$
\begin{align}
&\frac{dA}{dt}=\frac{1}{2}\varepsilon A\left( 1-\frac{1}{4}A^{2} \right)
\label{eqR109} \\
&\frac{d\theta}{dt}=-\varepsilon^{2}\frac{1}{4A}(2F_{1}(A)-F_{2}(A))
=-\varepsilon^{2}\frac{1}{8}\left( \frac{7}{32}A^{4}-A^{2}+1 \right)
\label{eqR110}
\end{align}

Equation \eqref{eqR109} has two fixed points, namely $A=0$ and $2$. The amplitude $A$ asymptotically approaches $A=2$, which is the limit cycle. 
Therefore, the dynamical behavior of $\theta (t)$ on the limit cycle reads
\begin{equation}
\theta(t) = -\varepsilon^{2}\frac{1}{16} t +\theta_{0},
\label{eqR111}
\end{equation}
where $\theta_{0}$ is the initial phase at $t=0$. Thus, the globally valid solution on the limit cycle is given by

\begin{equation}
x(t)=2\cos(\omega t+\theta_{0})
-\varepsilon\frac{1}{4}\sin(3\omega t+3\theta_{0})
-\varepsilon^{2}\frac{3}{32}\cos(3\omega t+3\theta_{0})
-\varepsilon^{2}\frac{5}{96}\cos(5\omega t+5\theta_{0})
+o(\varepsilon^{2}),
\label{eqR112}
\end{equation}

where the angular frequency $\omega$ up to the second order is expressed as
\begin{equation}
\omega = 1-\varepsilon^{2}\frac{1}{16}+o(\varepsilon^{2}).
\label{eqR113}
\end{equation}

Using Eqs. \eqref{eqR112} and \eqref{eqR113}, the period and $NS$ of the van der Pol model read

\begin{align}
&\tau =\frac{2\pi}{\omega}=\frac{2\pi}{1-\varepsilon^{2}/16+o(\varepsilon^{2})}
= 2\pi \left( 1+\varepsilon^{2}\frac{1}{16} +o(\varepsilon^{2}) \right) ,
\label{eqR114} \\
&NS=\left[ \frac{\sum_{j=1}^{\infty}|a_{j}|^{2}j^{2}}{\sum_{j=1}^{\infty}|a_{j}|^{2}} \right]^{\frac{1}{2}}
= \left[ \frac{1+\varepsilon^{2}(9/64)+o(\varepsilon^{2})}{1+\varepsilon^{2}/64+o(\varepsilon^{2})} \right]^{\frac{1}{2}}
= 1+\varepsilon^{2}\frac{1}{16}+o(\varepsilon^{2}),
\label{eqR115}
\end{align}

Accordingly, 

\begin{equation}
\tau =2\pi NS.
\label{eqR116}
\end{equation}

\begin{figure}[t]
 \begin{center}
  \includegraphics[width=10cm]{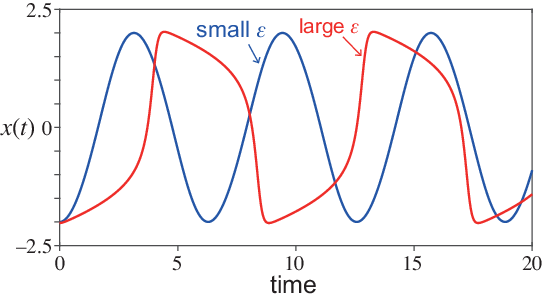}	
  \caption{The waveform examples of the van der Pol model. The parameter value is $\varepsilon =0.1$ (blue) and $\varepsilon =3$ (red).}	
 \label{figure8}
 \end{center}
\end{figure}

\begin{figure}[H]
 \begin{center}
  \includegraphics[width=16cm]{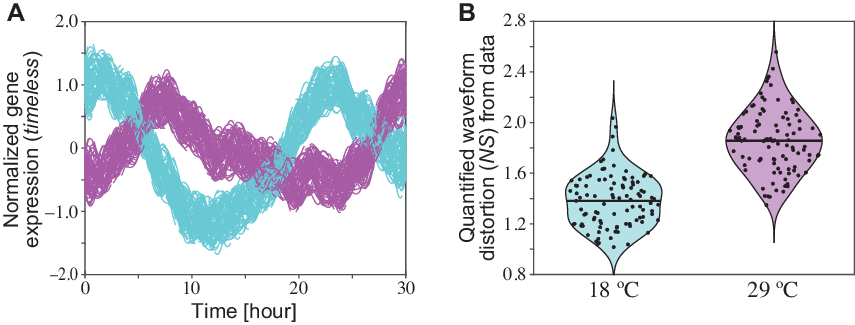}	
   \caption{Analysis of the circadian waveform of the Drosophila mutant perL 
   at different temperatures using previously reported experimental data 
   \cite{kidd15}. (A) Re-plot of Fig. 4B in \cite{kidd15}. 
   The average curves of \textit{tim-luc} at $18$ and $29$°C 
   are extracted using
   WebPlotDigitizer at 1-h intervals. Noise uniformly distributed between
   $-0.4$ and $0.4$ 
   is added to generate $100$ time series datasets. 
   Spline interpolation 
   is applied to set the sampling interval to $0.1$ h. 
   The interpolated time series data at $18$ (cyan) and $29$ ${}^\circ$C (magenta) 
   are plotted. (B) Distribution of the quantified waveform distortion ($NS$) 
   from the data at $18$ (cyan) and $29$ ${}^\circ$C (magenta). The noisy time series data 
   are detrended by 
   multiplying 
   an exponential 
   function to align local maxima at $18$ ${}^\circ$C or local minima at $29$ ${}^\circ$C. 
   The Fourier coefficients of the detrended time-series 
   are quantified using GHA. 
   $NS$ values were estimated from the coefficients 
   up to the third harmonics. Dots represent 
   the $NS$ values for each data set, and horizontal lines represent 
   their average values.}
 \label{figure1}	
 \end{center}
\end{figure}

This expression clearly illustrates that the period $\tau$ and the waveform distortion $NS$ in the van der Pol model tend to increase (or decrease) together in a proportional manner, which was also demonstrated using signal processing methods (Supplementary Fig. \ref{figure8}) \cite{kuro19}. 

\section*{Supplementary Figures}

\renewcommand{\tablename}{Supplementary Table }
\section*{Supplementary Tables}

\begin{table}[H]
 \caption{Activation energies and frequency factors for each reaction in Fig. \ref{figure3}}
 \label{tabS1}
 \begin{center}
  \begin{tabular}[H]{ccc} \hline
  Parameter & Activation Energy, $E_{i}$ & Frequency Factor, $A_{i}$ \\ \hline
  $k_{1}$ & $1.39 \times 10^{4}$ & $40.0$ \\
  $k_{2}$ & $6.31 \times 10^{3}$ & $2.31$ \\
  $k_{3}$ & $2.89 \times 10^{3}$ & $0.515$ \\
  $p_{1}$ & $1.94 \times 10^{4}$ & $229$ \\
  $p_{2}$ & $7.03 \times 10^{3}$ & $2.44$ \\
  $r$     & $8.11 \times 10^{4}$ & $1.06 \times 10^{13}$ \\ \hline
  \end{tabular}
 \end{center}
\end{table}

\begin{table}[H]
 \caption{Parameter values for each reaction in Fig. \ref{figure4}B-C and Fig. \ref{figure2}.}
 \label{tabS2}
 \begin{center}
  \begin{tabular}[H]{c|cc|cc} \hline
  & \multicolumn{2}{c|}{Fig. \ref{figure4}B and Fig. \ref{figure2}} & \multicolumn{2}{c}{Fig. \ref{figure4}C} \\
  Parameter & slow & fast & slow & fast \\ \hline
  $k_{1}$ & $0.269$  & $0.296$  & $0.247$  & $0.278$ \\
  $k_{2}$ & $0.200$  & $0.221$  & $0.192$  & $0.213$ \\
  $k_{3}$ & $0.0817$ & $0.150$  & $0.0662$ & $0.112$ \\
  $p_{1}$ & $0.290$  & $0.328$  & $0.160$  & $0.219$ \\
  $p_{2}$ & $0.246$  & $0.376$  & $0.256$  & $0.312$ \\
  $r$     & $0.180$  & $0.226$  & $0.244$  & $0.314$ \\
  $n$     & $15$     & $15$     & $13$     & $13$ \\ \hline
  \end{tabular}
 \end{center}
\end{table}


















\end{document}